\documentclass[12pt,a4paper]{article}
\usepackage[dvipdfmx]{graphicx,color}
\usepackage{amsmath, amssymb}
\usepackage{bm}

\newcommand{\vct}[1]{{\bm #1}}

\newcommand{\mtrix}[3]{\langle \,#1\,|\,#2\,|\,#3\,\rangle}

\newcommand{\avr}[1]{\langle \,#1\,\rangle}

\begin{document}

\begin{center}
{\Large The mean square radius of the neutron distribution\\
and the skin thickness derived from electron scattering\\[8pt]}

 
\vspace{5mm} 
\noindent
Haruki Kurasawa\footnote[1]{kt.suzuki2th@gmail.com}$^1$, Toshimi Suda$^2$
 and Toshio Suzuki$^2$\\ 
$^1$Department of Physics, Graduate School of Science, Chiba University \\Chiba 263-8522, Japan\\
$^2$Research Center for Electron Photon Science, Tohoku University \\Sendai 982-0826, Japan\\
\end{center}

\vspace{5mm}
\noindent
Abstract : The second-order moment of the nuclear charge density($R^2_c$)
is dominated by the mean square radius(msr) of the point proton
distribution($R_p^2$), while the fourth-order moment($Q^4_c$) depends on 
the msr of the point neutron one($R_n^2$) also.
Moreover, $R^2_n$ is strongly correlated to $R^2_c$ in nuclear models.
According to these facts, the linear relationship between various moments in
the nuclear mean field models are investigated with use of 
the least squares method for  $^{40}$Ca, $^{48}$Ca and $^{208}$Pb.
From the intersection points of the obtained straight lines
with those of the experimental values for $R^2_c$ and $Q^4_c$ determined through electron scattering,
the values of $R_p$ and $R_n$ are estimated.
Since relativistic and non-relativistic models provide different lines,
the obtained values of $R_n$ and the skin thickness($R_n-R_p$)
differ from each other in the two frameworks.

\section{Introduction}\label{intro}

It is a long standing problem how the neutrons are distributed in nuclei.
Even though it is one of the most fundamental problem in nuclear physics,
the neutron distribution is not well determined yet, since there is no simple
and reliable way to explore it experimentally\cite{thi}.

In contrast to the neutron distribution, the proton one is widely investigated
through the nuclear charge density observed by electron scattering\cite{vries}.
Electron scattering is an unambiguous tool to examine the nuclear
charge distribution, since the electromagnetic interaction and the reaction
mechanism are well understood theoretically\cite{deforest,bd}.

Recently, the interest in the neutron distribution in nuclei is rapidly
increased not only in nuclear physics, but also in other fields\cite{hargen}.
In nuclear physics, recent progress in the study
of unstable nuclei is expected to be more accelerated with further knowledge
of the excess neutron distribution\cite{suda}.
In astrophysics it is pointed out that the neutron distribution in nuclei
provides crucial information on the fundamental quantities, for example,
in the study of the neutron star. The difference between the mean square
radii of the proton and neutron distribution in nuclei has been shown
to be related to the problem on the size of the neutron star\cite{hargen}.

For the last ten years, responding to the above interest,
there have been noticeable development in the study of the neutron
distribution both experimentally and theoretically.
Experimentally, 
the parity-violating asymmetry ${\rm A_{pv}}$ in the polarized-electron
scattering has been measured\cite{abra}, which provides us with the ratio
of the weak charge form factor to the electromagnetic charge form factor. 
In the plane wave approximation, the latter is given by the Fourier
transform of the charge density, while the former by that of the weak
charge density for which the neutrons are  mainly responsible. 
The reaction mechanism of the parity-violating electron scattering is well
known and the analysis is similar to the conventional
electron scattering\cite{abra}.
Because of the weak process, however, experiment is much more difficult
and time consuming, compared with the conventional one. Indeed, 
the value of the form factor is available at present, only for $^{208}$Pb,
and at a single value of the momentum transfer, $q = 0.475 {\rm fm}^{-1}$
with the error of about 10\% due to the systematic
and statistic one\cite{abra}.
It is apparently impossible to determine the root mean square radius(rms)
of the point neutron distribution($R_n$) with a single value
from the experiment.

Nevertheless, the parity-violating electron scattering has 
brought a new insight in the study of the neutron distribution
by combining with the recent analysis of A$_{\rm pv}$ based on the 
nuclear mean field models which have been accumulated for several decades.
Using the 47 types of the nuclear Hamiltonian,
which reproduce well the gross properties of nuclei such as the binding
energies and the charge radii along the periodic table,
Roca-Maza et al. have shown that most of ${\rm A_{pv}}$ predicted by those
nuclear phenomenological models are on the straight regression line
as a function of $R_n$, as $10^7{\rm A_{pv}}=25.83-3.31R_n$ \cite{roca}.
This fact implies that if the experimental error is negligible,
the single value of ${\rm A_{pv}}$ is enough to estimate the value of $R_n$
which is expected in the mean field models.

Unfortunately, the above experimental error of ${\rm A_{pv}}$ is not
small enough to fix the value of $R_n$. The observed value yields $R_n$ in $^{208}$Pb
to be between 5.60 and 5.94 fm\cite{abra}, while the calculated line of $R_n$ spans
a narrower range from 5.55 to 5.80 fm\cite{roca}.
Although it may be difficult to determine the values of $R_n$
in the parity-violating electron scattering without the help of nuclear models,
more precise experiment is strongly desired. 
Both nuclear physics and astrophysics require less than 1\% accuracy
of the values of $R_n$ for their purposes\cite{thi}. 
Indeed, new experiment is planed aiming a small experimental error
in $^{48}$Ca and $^{208}$Pb\cite{thi}.

The purpose of the present paper is to estimate the values of $R^2_n$,
together with those of the mean square radius(msr) of the point proton
density($R^2_p$) and the skin thickness($\delta R=R_n-R_p$)
in the relativistic and non-relativistic mean field models,
by using the same method as for ${\rm A_{pv}}$\cite{abra},
but by employing the experimental data on the mean fourth-order moment($Q^4_c$)
together with  those on the msr($R^2_c$) of the nuclear charge density
observed through the conventional electron scattering\cite{vries,emrich}.

Recently, it has been shown that the msr of the nuclear charge density($R^2_c$)
is dominated by $R^2_p$, while the mean fourth-order moment of the charge density($Q^4_c$)
depends on $R^2_n$ also\cite{ksmsr}.
Moreover, $R^2_n$ is known to be strongly correlated with $R^2_c$ through
$R^2_p$ in nuclear models.
If the relationship between various calculated moments predict straight lines,
their intersection points with the lines for experimental values determine the values of
$R_p$ and $R_n$ in the framework of the mean field approximation.
The obtained values of $R_n$ are expected to be within a narrower range,
since the experimental errors are much smaller in the conventional electron
scattering\cite{vries} than in the parity-violating one in Ref.\cite{abra}.
As far as the authors know, this is the first paper to analyze the neutron density
distribution based on the experimental data from conventional electron scattering
in a long history of nuclear physics. 

For the present purpose, it is necessary to define exactly $R_c$ and $Q_c$
in both relativistic and non-relativistic ways.
In the following section, we briefly review the definition of 
$R_c$ and $Q_c$, according to Ref.$\cite{ksmsr}$.
Since the non-relativistic expression of $Q_c$ has not been discussed so far,
it will be derived with use of the Foldy-Wouthuysen(F-W) transformation
of the four-component framework to the two-component one,
in the same way as for $R_c$.

In order to show that the neutron density contributes appreciably to $Q_c$, 
$^{48}$Ca and  $^{208}$Pb will be taken, as examples, from nuclei for which
the experimental data are available at present\cite{vries, emrich}.
The moments of $^{40}$Ca will also be explored in detail in order to make clear
a role of the excess neutrons in $^{48}$Ca.
In \S\ref{example} will be shown the moments
of those nuclei calculated with a few relativistic and non-relativistic models,
before carrying out the least squares analysis. 
The structure of each moment will be seen numerically in detail.  
In \S\ref{reg}, the relationship between the various moments will be
analyzed, using 11 relativistic and 9 non-relativistic models which are chosen
arbitrarily from the literature.

It will be shown that
the relativistic and non-relativistic models yield different
linear relationships between moments from each other,
reflecting their different structures. As a result, the obtained values
of $R_p$ and $R_n$ are different in the two frameworks. 
On the one hand, relativistic models predict
the values of $R_n$ to be 3.587$\sim$3.605 fm for $^{48}$Ca
and 5.723$\sim$5.749 fm for $^{208}$Pb. Furthermore, the same analyses
determine the value of $R_p$, which yields the skin thickness
$\delta R=R_n-R_p$ to be 0.206$\sim$0.232 fm for $^{48}$Ca
and 0.258$\sim$0.306 fm for $^{208}$Pb.
On the other hand, non-relativistic models will provide the values of $R_n$
to be 3.492$\sim$3.502 fm for $^{48}$Ca and 5.587$\sim$5.627 for $^{208}$Pb,
together with the values of $\delta R$ to be 0.115$\sim$0.139
and 0.128$\sim$0.194 fm for $^{48}$Ca and $^{208}$Pb, respectively.
Thus, the values of $R_n$  and $\delta R$ from the non-relativistic models
is smaller about 0.1 fm in  both $^{48}$Ca and $^{208}$Pb
than those from the relativistic models.

In the above values, their ranges stem from the experimental errors.
The deviation from the mean value is less than $\pm 0.5\%$.
For example, in the case of $R_n$ in $^{208}$Pb, it is given as $\pm 0.227\%$
in the relativistic models, which is much smaller than in the previous study
in the parity-violating electron scattering\cite{abra}.
In the regression analysis, the confidence and prediction
bands may also be explored in addition to the least squares fitting,
as in ref.\cite{roca}.
It is not clear for the present authors, however,
whether or not a hypothesis of the normal probability
distribution holds with respect to the errors between the calculated values and
the fitting curves. Hence, as a measure of the theoretical errors,
the values of the standard deviation of the least square line will be provided.
In taking into account the standard deviation in addition to the experimental error,
the estimated range of the mean value, for example, of $R_n$ will be at most
$\pm 1\%$. 

The final section will be devoted to a brief summary. The structure of the least
squares analysis will be summarized in Appendix.

\section{The moment of the nuclear charge density}

We briefly review the definition of the mean 2nd-order moment($R^2_c$) and
the mean fourth-order moment($Q^4_c$) of the nuclear charge density \cite{ksmsr}
which is determined through electron scattering\cite{deforest, bd}.

In neglecting the center-of-mass correction,
the relativistic charge density of the nuclear ground state 
is given by\cite{ks}
\begin{equation}
 \rho_c(r) = \int \frac{d^3q}{(2\pi)^3}\exp(-i\vct{q}\!\cdot\!\vct{r})
 \tilde{\rho}(q).\label{rcd}
\end{equation}
Its Fourier component is described as\cite{ksmsr}
\begin{equation}
\tilde{\rho}(q) =\int d^3x\exp(i\vct{q}\!\cdot\!\vct{x})
 \sum_\tau\Bigl(G_{E\tau}(q^2)\rho_\tau(x)+F_{2\tau}(q^2)W_\tau(x)\Bigr),
 \label{fourier}
\end{equation}
where $G_{E\tau}(q^2)$ stands for the Sachs form factor,
$F_{2\tau}(q^2)$ the Pauli form factor\cite{bd}, and
$\tau$ represents proton($p$) and neutron($n$).
The {\it point} nucleon density $\rho_\tau$
and the spin-orbit density $W_\tau$ are given by\cite{ks}
\begin{align}
\rho_\tau(r)&= \mtrix{0}{\sum_{k\in\tau} \delta(\vct{r}-\vct{r}_k)}{0},\\
W_\tau(r)&= \frac{\mu_\tau}{2M}\left(-\frac{1}{2M}\vct{\nabla}^2\rho_\tau(r)
+i\vct{\nabla}\!\cdot\! \mtrix{0}{\sum_{k\in\tau}
\delta(\vct{r}-\vct{r}_k)\vct{\gamma}_k}{0}\right),
\end{align}
where $|\,0\,\rangle$ stands for the nuclear ground state,
and the subscript $k$ indicates the nucleon from 1 to $Z$ for $\tau=p$
and to $N$ for $\tau=n$.
Moreover, $M$ denotes the nucleon mass whose value will be
mentioned later, and $\mu_\tau$ the anomalous
magnetic moment to be $\mu_\tau=1.793$ for $p$ and $-1.913$ for $n$. 
The first equation satisfies $\int d^3r \, \rho_\tau (r) = Z$ for $\tau = p$
and $N$ for $\tau = n$, respectively, while the second equation 
$\int d^3r\, W_\tau(r)=0$, as it should.
Their explicit forms
in the relativistic  nuclear mean field models are written as\cite{ksmsr,ks}
\begin{align} 
\rho_\tau(r)&= \sum_{\alpha\in\tau} \frac{2j_\alpha+1}{4\pi r^2}
 \left(G_\alpha^2 + F_\alpha^2\right),\\
W_\tau(r)&= \frac{\mu_\tau}{M}\sum_{\alpha\in\tau} \frac{2j_\alpha+1}{4\pi r^2}
 \frac{d}{dr}\left(\frac{M-M^*(r)}{M}G_\alpha F_\alpha
+ \frac{\kappa_\alpha +1}{2Mr}G_\alpha^2 -
\frac{\kappa_\alpha - 1}{2Mr}F_\alpha^2\right).
\label{so}
\end{align}
In the above equations, $j_\alpha$ denotes the total angular momentum
of a single-particle, $\kappa_\alpha=(-1)^{j_\alpha-\ell_\alpha
 +1/2}(j_\alpha+1/2)$, $\ell_\alpha$ being the orbital angular momentum,
and $M^*(r)$ the nucleon effective mass defined
by $M^*(r)=M+V_\sigma(r)$, where $V_\sigma(r)$ represents the $\sigma$
meson-exchange potential which behaves in the same way
as the nucleon mass in the equation of motion.
The function 
$G_\alpha(r)$ and $F_\alpha(r)$ stand for the radial parts of
the large and small components of the single-particle wave function,
respectively, with the normalization,
\begin{equation}
\int_0^\infty\! dr \left(G_\alpha^2 + F_\alpha^2\right)=1.\label{norm}
\end{equation}
The spin-orbit density is a relativistic correction due to the anomalous
magnetic moment of the nucleon, and its role is enhanced by
the effective mass in relativistic nuclear models as seen
in Eq.(\ref{so})\cite{ks}.
The reason why Eq.(\ref{so}) is called the spin-orbit density
will be found in Refs.\cite{ksmsr, ks}.

The relativistic nuclear charge density Eq.(\ref{rcd}) is finally
written as,
\begin{equation}
 \rho_c(r) =\sum_\tau\Bigl( \rho_{c\tau}(r) + W_{c\tau}(r) \Bigr)\label{cd2}
\end{equation}
by convoluting a single-proton and -neutron density, 
\begin{align}
\rho_{c\tau}(r)&= 
\frac{1}{r}\int_0^\infty dx\, x\rho_\tau(x)\Bigl( g_\tau(|r-x|)
-g_\tau(r+x)\Bigr), \\[4pt]
W_{c\tau}(r)&= \frac{1}{r}\int_0^\infty dx\,x W_\tau(x)
\Bigl( f_{2\tau}(|r-x|)-f_{2\tau}(r+x)\Bigr),
\end{align}
with the functions,
\begin{equation}
g_\tau(x)= \frac{1}{2\pi}\int_{-\infty}^\infty dq\, e^{iqx}G_{E\tau}(q^2),
 \hspace{1cm}
f_{2\tau}(x)= \frac{1}{2\pi}\int_{-\infty}^\infty dq\, e^{iqx}F_{2\tau}(q^2).
\end{equation}

The momentum-transfer dependence of the nucleon form factors
is still under discussions theoretically\cite{sachs,licht,gmiller}.
Experimentally also there are various functional forms to fit
the electron scattering data at present\cite{kelly,kelly2}.
In the previous paper\cite{ksmsr}, the following Sachs and Pauli form factors 
was employed according to Refs.\cite{ks,bertozzi,eden,mey,plat},
\begin{align}
G_{Ep}(q^2)&= \frac{1}{(1+r_p^2q^2/12)^2},
 \qquad F_{2p}=\frac{G_{Ep}(q^2)}
 {1+q^2/4M^2}, \label{expff}\\[4pt]
G_{En}(q^2)&= \frac{1}{(1+r_+^2q^2/12)^2}- \frac{1}{(1+r_-^2q^2/12)^2},\qquad
 F_{2n}=\frac{G_{Ep}(q^2)-G_{En}(q^2)/\mu_n}{1+q^2/4M^2},\nonumber
\end{align}
with
\begin{equation}
 r_p=0.81\, \textrm{fm}, \qquad r_{\pm}^2=(0.9)^2
  \mp0.06 \, \textrm{fm}^2. 
  \label{oldr}
\end{equation}
In the present paper, we take the values used in Ref.\cite{rhoro}
\begin{equation}
  r_p=0.877\, \textrm{fm}, \qquad r_{\pm}^2=(0.830)^2
  \mp0.058 \, \textrm{fm}^2.\label{newr}
\end{equation}
In Ref.\cite{rhoro}, $G_{En}(q^2)$ is given by the form
\begin{equation}
 G_{En}(q^2)=-\frac{r_n^2q^2/6}{1+q^2/M^2}\frac{1}{(1+r_p^2q^2/12)^2},
\end{equation}
with $r_n^2=-0.116\, {\rm fm}^2$. This is numerically almost equal to
$G_{En}(q^2)$ in Eq.(\ref{expff}) with the values of Eq.(\ref{newr}),
and the values of the first and the second derivative of these form factors
are taken to be equal to each other at $q^2=0$.

There are still discussions on the values of $r_p$ and $ r_{\pm}^2$
themselves\cite{sick,data1,data2,xiong}.
Effects of the ambiguity on the nucleon size on the nuclear moments
will be seen later, in comparing the previous results\cite{ksmsr} with
the present ones.
The value $r_p=0.877$ fm is almost equal to the upper bound
$r_p=0.887$ fm of the proton size at present\cite{sick}. 

The relativistic charge density Eq.(\ref{cd2}) satisfies
$\int d^3r \, \rho_c(r) = Z$.
Then, the mean $2n$th-order moment $\avr{r^{2n}}_c$ of the nuclear
charge distribution is given by
\begin{equation}
\avr{r^{2n}}_c =\sum_\tau \avr{r^{2n}}_{c\tau},
\qquad
Z\avr{r^{2n}}_{c\tau}=\int d^3r\, r^{2n}\left(\rho_{c\tau}(r)
					  + W_{c\tau}(r) \right).\label{nth1}
\end{equation}
In calculating $\avr{r^{2n}}_c$, it is convenient to use the following
identity instead of the above equation itself \cite{ksmsr},
\begin{equation}
Z\avr{r^{2n}}_{c\tau}
 =(-\vct{\nabla}_q^2)^n \tilde{\rho}_\tau(q)|_{q=0}.\label{nth2}
\end{equation}
In the right-hand side, we have defined, according to Eq.(\ref{fourier}),
\begin{equation}
\tilde{\rho}(q) = \sum_\tau\tilde{\rho}_\tau(q).
\end{equation}

The second-order moment of the nuclear charge density
is obtained as the sum of the msr of the proton charge density, $R^2_{cp}$,
and the negative msr of the neutron charge density, $-R^2_{cn}$, \cite{ksmsr},
\begin{align}
\avr{r^2}_c=R^2_c&=R^2_{cp}-R^2_{cn},\label{rmsr}\\
                 &R^2_{cp}=R^2_p+R^2_{{\rm w}_p}+ r_p^2,\quad
 R^2_{cn}=-R^2_{{\rm w}_n}\frac{N}{Z} -(r_+^2-r_-^2)\frac{N}{Z},\nonumber
\end{align}
using the notations,
\begin{equation}
R^2_\tau=\avr{r^2}_\tau, \quad R^2_{{\rm w}_\tau}=\avr{r^2}_{W_\tau}.
 \nonumber
\end{equation}
Here, the following abbreviations are employed,
\begin{equation}
\avr{r^n}_\tau = \frac{1}{N_\tau}\int d^3r\, r^n\rho_\tau(r),
\qquad
\avr{r^n}_{W_\tau}=\frac{1}{N_\tau}\int d^3r\, r^nW_\tau(r),\label{point}
\end{equation}
with $N_p=Z$ and $N_n=N$.
In Eq.(\ref{rmsr}), $R^2_p$ in $R^2_{cp}$ represents the msr of the
point proton density. 
The second term in $R^2_{cp}$ and the first term in $R^2_{cn}$
come from the spin-orbit densities of protons and neutrons,
respectively. The last terms in $R^2_{cp}$ and $R^2_{cn}$
are the contributions from a single-proton
and a single-neutron size, which are not negligible
in the present discussions, as mentioned later.
We note that $R^2_{cn}$ has been defined so as to be positive.

The mean fourth-order moment of the nuclear charge density($Q^4_c$)
is given by Eq.(\ref{nth1}) and (\ref{nth2}) in terms of the proton
and neutron contributions,
\begin{equation}
\avr{r^4}_c =Q^4_c= Q^4_{cp}-Q^4_{cn}\,,\label{4thm}
\end{equation} 
where we have defined
\begin{align*}
Q^4_{cp}&= Q_{p}^4+Q_{2p}+Q_{2W_p}+Q_{4W_p}+(Q_4)_p\,, \\
Q^4_{cn}&= Q_{2n}+Q_{2W_n}+Q_{4W_n}+(Q_4)_n\,,
\end{align*}
with the notations for the protons,
\begin{align*}
 Q_{p}^4&= \avr{r^4}_p,\qquad
  Q_{2p}=\frac{10}{3}r_p^2\avr{r^2}_p,\qquad
 Q_{2W_p}= \frac{10}{3}(r_p^2+\frac{3}{2M^2})\avr{r^2}_{W_p},\qquad\\
 Q_{4W_p}&= \avr{r^4}_{W_p},\qquad
(Q_4)_p =\frac{5}{2}r_p^4,
 \end{align*}
and for the neutrons, 
 \begin{align*}
 Q_{2n}&=-\frac{10}{3}(r_+^2-r_-^2) \avr{r^2}_n\frac{N}{Z},\qquad
Q_{2W_n}=-\frac{10}{3}(r_p^2+\frac{3}{2M^2}-\frac{r_+^2-r_-^2}
 {\mu_n})\avr{r^2}_{W_n}\frac{N}{Z},\qquad\\
 Q_{4W_n}&=-\avr{r^4}_{W_n}\frac{N}{Z},\qquad
(Q_4)_n =-\frac{5}{2}(r_+^4-r_-^4) \frac{N}{Z}. 
 \end{align*}
The details of the derivation will be found in Ref.\cite{ksmsr}.
The number of the components is increased, compared with the one of $R^2_c$,
but the meaning of each terms may be clear.
It should be noticed that while $R_c$ is independent of the point
neutron density as in Eq.(\ref{rmsr}),
$Q_c$ depends on it through its msr in $Q_{2n}$.

Eq.(\ref{rmsr}) and (\ref{4thm}) should be used
within a relativistic framework.
In non-relativistic models, we need 
the expressions of the msr and mean fourth-order moment which are  equivalent
to the above equations up to $1/M^2$. It is obtained according to the 
F-W unitary transformation of the four-component
framework to the two-component one.
The F-W transformation for Dirac equation with electromagnetic field
has been performed by various authors\cite{bertozzi,macvoy,nishizaki}.
In the case of the relativistic Hamiltonian in the $\sigma$-$\omega$ model,
Nishizaki, et al.\cite{nishizaki} have obtained the
charge operator $\hat{\rho}(q)$ for
$\tilde{\rho}(q)=\mtrix{0}{\hat{\rho}(q)}{0}_{\rm{nr}}$ 
up to order $1/M^{*2}(r)$. Here, the matrix element, as indicated by the
subscript nr, is calculated using the wave functions in the
two component framework, and the operator is written as\cite{ksmsr} 
\begin{align}
\hat{\rho}(q) = \sum_{k=1}^Ae^{i\vct{q}\cdot\vct{r_k}}\Bigl( D_{1k}(q^2)
 + iD_{2k}(q^2)\vct{q}\!\cdot\!(\,\vct{p}_k\times\vct{\sigma}_k\,)\Bigr),
\label{fw}
\end{align}
where $D_1$ and $D_2$ are defined as
\begin{align}
 D_{1k}(q^2)&= F_{1k}(q^2)-\frac{q^2}{2}D_{2k}(q^2), \label{d0m}\\[4pt]
 D_{2k}(q^2)&= \frac{1}{4M^{*2}(r_k)}\left(F_{1k}(q^2)+
 2\mu_k F_{2k}\frac{M^*(r_k)} {M}\right),\label{d2m}
\end{align}
with the Dirac form factor $F_{1}(q^2)$ related to the Sachs and Pauli form
factor as\cite{bd}
\begin{equation}
F_{1\tau}(q^2)=G_{E\tau}(q^2)+\mu_\tau q^2F_{2\tau}(q^2)/(4M^2).\label{sachs}
\end{equation}
Then, using the equation
\begin{equation}
 -\vct{\nabla}^2_{\vct{q}}\hat{\rho}(q)|_{\vct{q}=0}
  =\sum^A_{k=1}\left(F_{1k}(0)r^2_k+(2\vct{\ell}_k\cdot\vct{\sigma}_k+3)
	       D_{2k}(0)-6F'_{1k}(0)\right),  \nonumber
\end{equation}
Eq.(\ref{nth2}) provides the non-relativistic expression for the msr
of the nuclear charge density,
\begin{equation}
R_{c,{\rm nr}}^2
 =\frac{1}{Z}\mtrix{0}{\sum_{k=1}^Z r_k^2}{0}_{\mathrm{nr}}
 -6G'_{E_p}(0)-6G'_{E_n}(0)\frac{N}{Z}+C_{\rm rel}. \label{correct}
\end{equation}
Here, $C_{\rm rel}$ represents the relativistic correction up
to order of $1/MM^\ast(r)$ and $1/M^{\ast 2}(r)$, which is described as
\begin{align}
 C_{\rm rel}=\mtrix{0}{\frac{1}{2Z}\sum_{k=1}^A\frac{\mu_k
 \left(2\vct{\ell}_k\!\cdot\!\vct{\sigma}_k + 3(1-M^\ast(r_k)/M)\right)}{MM^\ast(r_k)}
 +\frac{1}{4Z}\sum_{k=1}^Z \frac{2\vct{\ell}_k\!\cdot\!\vct{\sigma}_k +3}
 {M^{\ast 2}(r_k)}}{0}_{\mathrm{nr}}.\label{rr}
\end{align}

When using the free Dirac equation for the Hamiltonian,
the above relativistic correction is reduced to\cite{ksmsr}
\begin{equation}
C_{\rm rel}= \frac{1}{M^2}\left(\frac{1}{Z}\sum_{k=1}^A\mu_k
 \mtrix{0}{\vct{\ell}_k\!\cdot\!\vct{\sigma}_k}{0}_{\mathrm{nr}}+\frac{3}{4}
 +\frac{1}{2Z}\sum_{k=1}^Z\mtrix{0}
 {\vct{\ell}_k\!\cdot\!\vct{\sigma}_k}{0}_{\mathrm{nr}} \right).\label{r}
\end{equation}
It is convenient for the expression of $R_{c,{\rm nr}}^2$ to define 
\begin{equation}
\avr{r^n}_{\tau,{\rm nr+r}}
=\avr{r^n}_{\tau,\,{\rm nr}}+\frac{n(n+1)}{8M^2}
 \avr{r^{n-2}}_{\tau,\,{\rm nr}}+\frac{1}{2\mu_\tau}\avr{r^n}_{W_\tau,\,{\rm nr}},
 \label{equivalent}
\end{equation}
with
\begin{equation}
\avr{r^n}_{\tau,{\rm nr}}
=\frac{1}{N_\tau}
\mtrix{0}{\sum_{k\in\tau}r_k^{n}}{0}_{\rm nr}, \quad
 \avr{r^n}_{W_\tau,{\rm nr}}=\frac{n}{N_\tau}\frac{\mu_\tau}{2M^2}
 \mtrix{0}{\sum_{k\in\tau}r_k^{n-2}\vct{\ell}_k\cdot\vct{\sigma}_k}{0}_{\rm nr}.
 \label{wso}
\end{equation}
Then, $R_{c,{\rm nr}}^2$ for $M^\ast=M$ is expressed as
\begin{equation}
R_{c,{\rm nr}}^2
 =\avr{r^2}_{p,{\rm nr+r}}+\avr{r^2}_{W_p,{\rm nr}}-6G'_{E_p}(0)
+\avr{r^2}_{W_n,{\rm nr}}\frac{N}{Z}-6G'_{E_n}(0)\frac{N}{Z},\label{nmsr}
\end{equation}
which is a similar form to the relativistic one in Eq.(\ref{rmsr}).
It should be noticed that the first term of Eq.(\ref{nmsr}) is not the msr of the
point proton density, but including the relativistic corrections.
In order to make clear the difference between the msr of the point nucleon density
in the relativistic and non-relativistic models, the following notations for no-relativistic
models will be used,
\begin{equation}
 R^2_{p,{\rm nr}}=\avr{r^2}_{p,{\rm nr}},\qquad
  R^2_{n,{\rm nr}}=\avr{r^2}_{n,{\rm nr}}.\label{notation}
\end{equation}

We note that the terms of the right-hand side in Eq.(\ref{correct})
are formally consistent with each other up to order $1/M^{\ast 2}$ or $1/M^2$,
but that at present the values of $G'_{E_\tau}(0)$ are unknown
theoretically\cite{sachs,licht,gmiller}.
In the relativistic expression Eq.(\ref{rmsr}), they are taken
from Eq.(\ref{expff}) determined by experiment with use of the relationship,
\begin{equation}
 6G'_{E_p}(0)=-r_p^2\, , \hspace{5mm}   6G'_{E_n}(0)=-(r_+^2-r_-^2)\,.
\label{pn}
\end{equation}
If the same values are employed in Eq.(\ref{correct}),
the consistency in the non-relativistic expression becomes obscure.
This ambiguity is unavoidable at present,
although in the difference between two msr's like the isotope shift,   
the contribution from
the proton's form factor disappears and from the neutron's one
is reduced.

More strictly speaking, it is not possible to obtain the relativistic
corrections which are consistent with the non-relativistic mean field
models widely used at present, 
since their original four-component models are not known.
Because of this fact, the previous papers of
the non-relativistic models are forced to use a part of Eq.(\ref{r})
for the free Dirac Hamiltonian\cite{sly}. 
As a result, some parts of the relativistic corrections may be
included in the first term of Eq.(\ref{correct}) calculated in the
non-relativistic models
where the experimental values of $R_c$ are employed
as a input for fixing free parameters of nuclear interactions.
This kind of the inconsistency is a common
problem in discussing relativistic corrections to non-relativistic
models, in spite of the fact that those corrections should be there\cite{ksg}.

There may be two extreme standpoints in discussing relativistic corrections.
The one is that $R_c^2$ calculated so as to reproduce the experimental values
in non-relativistic phenomenological models
implicitly includes all relativistic corrections.
The other is that, without taking care of the inconsistency strictly,
all relativistic corrections are added to the first term of Eq.(\ref{correct}).
In the present paper, we will take the latter position
for calculations of $R_c^2$, using  Eq.(\ref{r}) together with (\ref{pn}).

The non-relativistic expression of the mean fourth-order
moment of the nuclear charge density, which is equivalent to Eq.(\ref{4thm}),
is derived in the same way as for the msr in Eq.(\ref{correct}). 
Eq.(\ref{fw}) provides up to $1/M_k^{\ast 2}$ and $1/MM^{\ast}$
\begin{align}
\vct{\nabla}^2_\vct{q}\vct{\nabla}^2_\vct{q}\hat{\rho}(q)|_{\vct{q}=0}
=\sum^A_{k=1}\left[\delta_{kp}\left(r^4_k+\frac{5r^2_k}{2M_k^{\ast 2}}
 +\frac{r_k^2\vct{\ell}_k\cdot\vct{\sigma}_k}{M_k^{\ast 2}}\right)
	     -20G'_k(0)\left(r^2_k+\frac{3}{4M_k^{\ast 2}}
+\frac{\vct{\ell}_k\cdot\vct{\sigma}_k}{2M_k^{\ast 2}}\right)\right.\nonumber\\
\left.+60G''_k(0)+\frac{2\mu_kr_k^2\vct{\ell}_k\cdot\vct{\sigma}_k}{MM_k^\ast}
-20F'_{2k}(0)\frac{\mu_k\vct{\ell}_k\cdot\vct{\sigma}_k}{MM_k^\ast}
+\left(1-\frac{M}{M_k^\ast}\right)\frac{5\mu_k}{M^2}\left(6F'_{2k}(0)
-r_k^2\right)\right],\nonumber
\end{align}
with the notation $M_k^\ast=M^\ast(r_k)$. According to Eq.(\ref{nth2}),
the ground-state expectation value of the above equation gives
the non-relativistic expression of the fourth-order moment, $Q^4_{c,{\rm nr}}$.
In the case of the Dirac Hamiltonian with $M_k^\ast=M$,
it is written as
\begin{equation}
Q^4_{c,{\rm nr}}=Q^4_{cp,{\rm nr}}-Q^4_{cn,{\rm nr}}\,,\label{n4thm}
\end{equation}
with
\begin{align*}
Q^4_{cp,{\rm nr}}&= Q_{p,{\rm nr+r}}^4+Q_{2p,{\rm nr+r}}+Q_{2W_p,{\rm nr}}
+Q_{4W_p,{\rm nr}}+(Q_4)_p\,, \\
Q^4_{cn,{\rm nr}}&= Q_{2n,{\rm nr+r}}+Q_{2W_n,{\rm nr}}+Q_{4W_n,{\rm nr}}+(Q_4)_n\,,
\end{align*}
which has the same form as the relativistic expression in Eq.(\ref{4thm}),
but by replacing $\avr{r^n}_\tau$ with $\avr{r^n}_{\tau,\,{\rm nr+r}}$,
and $\avr{r^n}_{W_\tau}$ with $\avr{r^n}_{W_\tau,\,{\rm nr}}$,
given in Eq.(\ref{equivalent}) and (\ref{wso}).

\section{Structure of the 2nd and the 4th-order moment}
\label{example}

Before the least squares analysis is performed in the following section,
it may be useful to understand the contribution of each component
of the second and the fourth-order moment numerically
by taking a few examples of nuclear models.

Table 1 shows the contribution of the components in Eq.(\ref{rmsr})
to $R^2_c$ and in Eq.(\ref{correct}) to $R^2_{c,{\rm nr}}$
in units of fm$^2$ for $^{40}$Ca, $^{48}$Ca and $^{208}$Pb.
They are calculated by employing three examples of the mean field models.
The two of them are relativistic nuclear models
named NL3\cite{nl3} and NL-SH\cite{nlsh}, while the rest is
the non-relativistic one SLy4\cite{sly4}. These are typical examples
of the nuclear models which have widely been used to explain nuclear structure
phenomenologically\cite{abra,roca,rhoro,nl3,nlsh,sly4}, 
and will also be used in the next section.

In the relativistic cases, the sum of $R^2_p$ and $r^2_p$
and each of the rest in Eq.({\ref{rmsr}}) are listed separately.
In non-relativistic calculations of Eq.(\ref{correct}),
its second and the third term are taken from Eq.(\ref{pn}), as in
the relativistic models.
The sum of  $R^2_{p,{\rm nr}}$ and $r^2_p$ is listed in the first column 
for that of $R^2_p$ and $r^2_p$.
As the relativistic corrections $C_{\rm rel}$, Eq.(\ref{r}) for the
Dirac Hamiltonian is used, since it is not able to derive
the corrections which are consistent with the non-relativistic
phenomenological models, as mentioned before. 
The values of the first term in Eq.(\ref{r}) are listed
as $\avr{r^2}_{W_p}$ and $\avr{r^2}_{W_n}N/Z$,
while the second term, $3/4M^2=0.0331\,\textrm{fm}^2$, is included
in $R^2_{c,{\rm nr}}$ listed in the column of $R^2_c$ in the Table 1.
The last term of $C_{\rm rel}$ in Eq.(\ref{r}) does not contribute to 
$R^2_{c,{\rm nr}}$ of Ca isotopes, but does to that of $^{208}$Pb. Its value,
0.0162 fm$^2$, is added to $R^2_{c,{\rm nr}}$ of $^{208}$Pb.  

The experimental values of the msr employed as inputs for fixing
the parameters of the nuclear models are also listed in Table 1,
according to the Refs.\cite{nl3,nlsh,sly4}, where
NL3 and NL-SH refer to Ref.\cite{vries}, while Sly4 to ref.\cite{otten}.
In the parentheses, the calculated values in the Refs.\cite{nl3,nlsh,sly4}
are shown for reference.

Table 2 shows the contribution of each term of Eq.(\ref{4thm}) to $Q^4_c$,
except for $Q_{2W_\tau}$ and $(Q_4)_\tau$. The contributions of $Q_{2W_p}$ and
 $Q_{2W_n}$ are listed together as $Q_{2W}=Q_{2W_p}-Q_{2W_n}$
 and those of $(Q_4)_\tau$ are
included in $Q_c^4$. The values of $((Q_4)_p-(Q_4)_n)$ are
given as, 1.0793, 0.9195, and 0.8649 fm$^4$ for $^{40}$Ca, $^{48}$Ca
and $^{208}$Pb, respectively.
In non-relativistic calculations with SLy4\cite{sly4},
Eq.(\ref{n4thm}) is used. The value of each term is listed in the same way as
for the relativistic one corresponding to it.

We note that in the calculation of the Coulomb energy, the only
direct term is taken into account in the relativistic models, while in the
non-relativistic models the exchange term also is evaluated as usual.
In the previous paper\cite{ksmsr}, in both relativistic and non-relativistic
models, the only direct term has been estimated.
In the present paper, the $J^2$ term\cite{sly4} of the spin-orbit potential in the
non-relativistic models is disregarded.

\begin{table}
\begin{tabular}{|l||r|r|r|r|r||r|} \hline
\rule{0pt}{12pt} & 
 $ R_p^2+r_p^2$  &
 $R^2_{W_p}$ &
 $R^2_{W_n}N/Z$&
 $(r_+^2 - r_-^2)N/Z$ &
\multicolumn{1}{c||}{$R_c^2$} &
\multicolumn{1}{c|}{Exp.} \\ \hline
\rule{0pt}{12pt}%
NL3        &          &           &          &           &          &          \\
$^{40}$Ca  & $12.173$ & $ 0.0222$ & $ -0.0244$ & $ -0.1160$ & $ 12.055$ & $11.90(12.03)$ \\
$^{48}$Ca  & $12.186$ & $ 0.0263$ & $ -0.1573$ & $ -0.1624$ & $ 11.892$ & $11.91(12.04)$ \\ 
$^{208}$Pb & $30.580$ & $ 0.1054$ & $ -0.1460$ & $ -0.1782$ & $ 30.362$ & $30.28(30.47)$  \\
\hline
\rule{0pt}{12pt}%
NL-SH      &          &           &          &           &          &          \\
$^{40}$Ca  & $ 12.041$ & $ 0.0225$ & $ -0.0248$ & $ -0.1160$ & $ 11.923$ & $11.90(11.88)$ \\
$^{48}$Ca  & $ 12.111$ & $ 0.0268$ & $ -0.1594$ & $ -0.1624$ & $ 11.816$ & $11.91(11.86)$ \\
$^{208}$Pb & $ 30.417$ & $  0.1071$ & $ -0.1482$ & $ -0.1782$ & $ 30.197$ & $30.28(30.22)$ \\
\hline
\rule{0pt}{12pt}%
SLy4       &          &           &          &           &          &          \\
$^{40}$Ca  & $12.463$ & $ 0.0000$ & $ 0.0000$ & $ -0.1160$ & $ 12.381$ & $12.18(12.20)$ \\
$^{48}$Ca  & $12.691$ & $ 0.0000$ & $ -0.1014$ & $ -0.1624$ & $ 12.461$ & $12.11(12.33)$ \\
$^{208}$Pb & $30.555$ & $ 0.0579$ & $ -0.0865$ & $ -0.1782$ & $ 30.397$ & $30.25(30.23)$ \\ \hline
\end{tabular}
\caption{
The mean square radius(msr) of the charge distribution
of $^{40}$Ca, $^{48}$Ca and $^{208}$Pb in units of fm$^2$.
The calculated values are listed, using parameters of the relativistic nuclear
models NL3\cite{nl3} and NL-SH\cite{nlsh}, and of the non-relativistic one SLy4\cite{sly4}.
The experimental values are those employed in the nuclear models to fix their parameters.
The evaluated values in Refs.\cite{nl3,nlsh,sly4} are listed
in the parentheses, respectively. For details, see the text.  
}
\end{table}

Experimental values in Table 2 are obtained using Fourier-Bessel analyses
with the data in Ref.\cite{vries}. In the next section,
we will use the data for Ca isotopes in Ref.\cite{emrich},
since it provides us with the experimental values of $Q_c$ together with
the experimental errors which play an essential
role in our purposes, as mentioned in \S\ref{intro}.
Refs.\cite{vries} and \cite{emrich} give the same values of $Q_c$,
up to the third digit after the decimal point,
which are the fourth-root of the listed experimental values.

Now a few comments should be mentioned.
The first is about the number of the digits of which we take
care in discussing $R_c$ and $Q_c$ in units of fm.
In the next section, we will round their values
off to three decimal places. There are two reasons. The one is
that the experimental errors of $R_c$ and $Q_c$ in Ref.\cite{emrich} are
$\pm$(0.009$\sim$0.022) fm, as will be shown in the next section.
The other is that in comparing the present tables with those of the previous
paper\cite{ksmsr}, it is seen that owing to the change of the nucleon size
from Eq.(\ref{oldr}) to Eq.(\ref{newr}),
for example in NL3, the calculated $R_c$ is increased by 0.017 fm
in Ca isotopes and 0.011 fm in $^{208}$Pb.
Thus, there is ambiguity experimentally on the values of $R_c$ and $Q_c$
at the second decimal place.
When we present the experimental and evaluated values of $R_c^2$
instead of $R_c$, and $Q_c^4$ instead of $Q_c$, we keep the numbers as in Table 2,
so as to reproduce the values of $R_c$ and $Q_c$ up to the third decimal place.

\begin{table}
\begin{tabular}{|l||r|r|r|r|r|r|r||r|} \hline
\rule{0pt}{12pt}
 & \multicolumn{1}{c|}{$Q_{p}^4$} & \multicolumn{1}{c|}{$Q_{2p}$} & \multicolumn{1}{c|}{$Q_{2n}$}
 & \multicolumn{1}{c|}{$Q_{2W}$} & \multicolumn{1}{c|}{$Q_{4W_p}$} & \multicolumn{1}{c|}{$Q_{4W_n}$}
 & \multicolumn{1}{c||}{$Q_c^4$}   &  \multicolumn{1}{c|}{Exp.} \\ \hline
\rule{0pt}{12pt}NL3    &   &  &  &  &  &  &  &  \\ 
$^{40}$Ca & $183.921$ & $ 29.238$ & $ 4.284$ & $-0.001$ & $0.457$ & $ 0.525$ & $209.884$ & $199.991$ \\
$^{48}$Ca & $178.085$ & $ 29.270$ & $ 7.035$ & $-0.333$ & $0.738$ & $ 4.962$ & $196.682$ & $194.714$ \\  
$^{208}$Pb& $1115.64$ & $ 76.429$ & $19.579$ & $-0.083$ & $8.269$ & $12.304$ & $1169.24$ & $1171.58$ \\
\hline
\rule{0pt}{12pt}NL-SH      &   &  &  &  &  &   &  & \\ 
$^{40}$Ca & $178.612$ & $ 28.899$ & $ 4.236$ & $-0.001$ & $0.499$ & $ 0.569$ & $204.282$ & $199.991$ \\ 
$^{48}$Ca & $174.602$ & $ 29.079$ & $ 6.950$ & $-0.337$ & $0.780$ & $ 4.973$ & $193.119$ & $194.714$ \\
$^{208}$Pb& $1098.69$ & $ 76.009$ & $19.378$ & $-0.085$ & $8.381$ & $12.427$ & $1152.06$ & $1171.58$ \\
\hline
\rule{0pt}{12pt}SLy4  &    &  &  &  &  &  &   & \\
$^{40}$Ca & $194.031$ & $ 30.066$ & $ 4.410$ & $-0.000$ & $-0.070$ & $-0.061$  & $220.757$ & $199.991$ \\
$^{48}$Ca & $196.484$ & $ 30.651$ & $ 7.068$ & $-0.262$ & $-0.002$ & $3.515$ & $217.206$ & $194.714$ \\
$^{208}$Pb& $1125.87$ & $ 76.489$ & $18.777$ & $-0.062$ & $ 4.693$ & $7.531$ & $1181.55$ & $1171.58$ \\
\hline
\end{tabular}
\caption{
The 4th order moment of the charge distribution
of $^{40}$Ca, $^{48}$Ca and $^{208}$Pb. 
The value of each term in Eq.(\ref{4thm}) is listed
in units of fm$^4$, but the values of $Q_{2W_p}$ and
 $Q_{2W_n}$ are given together as $Q_{2W}=Q_{2W_p}-Q_{2W_n}$
 and those of $((Q_4)_p-(Q_4)_n)$ are included in $Q_c^4$.
They are 1.0793, 0.9195, and 0.8649 fm$^4$ for $^{40}$Ca, $^{48}$Ca
and $^{208}$Pb, respectively.
The experimental values are obtained by the Fourier-Bessel analyses
of data in Refs.\cite{vries,emrich}. 
For details, see the text.
}
\end{table}

The second comment is that Table 1 shows $R^2_c$
to be dominated by $R^2_p(R^2_{p,{\rm nr}})$ with $r_p^2$.
The contributions of the rest, however, change
the values of the second digit after the decimal point in $R_c$.
Hence, we will include fully their contributions in the calculations
of the next section also.

The third comment is on $Q_c$ in Table 2. 
As the details have been discussed in Ref.\cite{ksmsr}, it shows that
the sum of $Q_{p}^4(Q^4_{p,{\rm nr+r}})$ and $Q_{2p}(Q_{2p,{\rm nr+r}})$
overestimates the experimental values.
Thus, it is necessary to have negative contributions from the neutron
density through $Q_{2n}(Q_{2n,{\rm nr+r}})$ and $Q_{4W_n}(Q_{4W_n,{\rm nr}})$.
In $^{48}$Ca, $Q_{2n}$ reduces the value of $Q_{2p}$
by about 24.0$\%$ in the case of NL3. 
The sum of $Q_{2n}$ and $Q_{4W_n}$ amounts to
40.0$\%$ of the sum of $Q_{2p}$ and $Q_{4W_p}$.
In $^{208}$Pb, $Q_{2n}$ reduces the value of $Q_{2p}$
by about 25.6$\%$ in the case of NL3,
and the sum of  $Q_{2n}$ and $Q_{4W_n}$ is 37.6\%
of the sum of $Q_{2p}$ and $Q_{4W_p}$.
The main term of $Q_c$ is $Q_{p}^4$,
to which the ratio of the sum of $Q_{2n}$ and $Q_{4W_n}$ is 6.74\% in
$^{48}$Ca and 2.86\% in $^{208}$Pb, in the case of NL3.
In spite of the fact that the number of the neutrons is larger in $^{208}$Pb
than in $^{48}$Ca, their contribution is decreased. This result is due to
the constraint on the A$^{2/3}$-dependence of the msr of the nuclear
matter density in the stable nuclei.
The contribution of the neutrons to $Q_c$ is thus not so large in stable nuclei,
but will  clearly be seen in the least squares analysis of the following section.

\section{The least squares analysis of the moments}\label{reg}

The previous section have provided us with understanding
how each component contributes to the moments, but 
the meaning of small change in the numbers from one model to another
is not obvious. 
All of the phenomenological models discussed in this paper employ
the experimental values of $R_c$ as inputs together with other
fundamental quantities like the binding energies and some nuclear matter
properties, in order to fix their free parameters of the interactions.
Among the inputs, a special attention is paid
for reproducing the values of $R_c$ \cite{nl3}.
Hence, except for some cases, the calculated values, in particular
within relativistic models, differ from one another only at the second digit
after the decimal point. These differences, however, do not seem to have
a special meaning, since each model is constructed according to different
inputs and to different aims to reproduce various nuclear
properties\cite{nl3,nlsh,sly4}.
Moreover, sometimes the input values of $R_c$ are different
among the models as in Table 1.
Therefore, it is better to find common constraints
on the values obtained by the phenomenological models
in reproducing the experimental data,
rather than to choose the one model with the best fit by
comparing the predicted values of each model with experiment.
One way to find such common constraints is to use the least squares analysis(LSA),
as employed in Refs.\cite{abra,roca} in order to
find the relationship between $R_n$ and ${\rm A_{pv}}$ in the mean field models.
We will follow their method to explore the msr of proton and neutron
distributions of $^{40}$Ca, $^{48}$Ca and $^{208}$Pb in the relativistic
and non-relativistic mean field models.

\subsection{The rms of the proton
  and neutron densities in $^{40}$Ca}\label{ca40}

Experimental values of $R_c$ and $Q_c$ are provided
in units of fm in Refs.\cite{vries,emrich}.
From Eq.(\ref{rmsr}) and (\ref{correct}), however, it may be more reasonable
to analyze the relationship between $R_c^2$(fm$^2$) and $R_p^2$(fm$^2$)
than that between $R_c$(fm)and $R_p$(fm). 
In the case of $Q_c$, Eq.(\ref{4thm}) and (\ref{n4thm}) give the relationship
between various moments in units of fm$^4$.
In the following LSA, therefore, we will compare all the moments with each other
in their own units in those expressions, for example, as  $Q_c^4$ (fm$^4$)
against $R_p^2$ (fm$^2$).
Then, the experimental values of $R_c^2$ and $Q_c^4$ including their errors
are expressed in such a way that the square and fourth root reproduce the
experimental ones of $R_c$ and $Q_c$ in Refs.\cite{vries, emrich}, respectively.

The  LSA will be performed between  $R^2_p$ and $R_c^2$,
between $R_p^2$ and $Q_c^4$, and between $R^2_p$ and $Q^4_{cp}$
for the proton density, and similarly between those for the neutron density replacing $p$
with $n$ in the above quantities.
The LSA is also performed between $Q^4_{cp}$ and $Q^4_c$ in order to separate $Q^4_{cp}$
from $Q^4_c$, and $Q^4_{cn}$ will be obtained through the definition $Q^4_{cn}=Q^4_{cp}-Q^4_c$,
after $Q^4_{cp}$ is fixed. The fixed value of $Q^4_{cp}$ will be used
as the {\it pseudo} experimental value.
In the non-relativistic models, all the above quantities are replaced with the corresponding
non-relativistic ones, $R^2_{p,{\rm,nr}}$, etc.
The notation of $Q^4_{p,{\rm nr}}$ for the fourth-order moment of the non-relativistic
point proton density will be used, which is given by $\avr{r^4}_{p,\,{\rm nr}}$
in Eq.(\ref{wso}), as for $R^2_{\tau,\,{\rm nr}}$, in order to distinguish it
from $Q^4_{p,{\rm nr+r}}=\avr{r^4}_{p,\,{\rm nr+r}}$ in Eq.(\ref{n4thm}).

The intersection point of the obtained least square line(LSL) with the line of the experimental
value of $R^2_c$ or $Q^4_c$ will determine the accepted range of the msr of the point proton
and that of the point neutron density. The above three kinds of the LSL yield the three accepted
ranges, of which the common range provides the final accepted range $\mathcal{R}$
in the mean field models. 

We note that, as in the regression analysis, it is not necessary for elements of the moment
to be independent of one another, or for the relationship between the moments to be described
explicitly in the present LSA.
The moment $R_c^2$ does not depend explicitly on $R_n^2$, but in the mean field models,
$R_n^2$ may be strongly constrained by $R_p^2$ which dominates $R_c^2$.
Hence, it is reasonable to expect a well defined LSL between $R_n^2$ and $R_c^2$.
It will be seen that the fitting line between $R_n^2$ and $R_c^2$ does not contradict other lines,
and in some cases, makes actually the accepted range of $R^2_n$ narrower.
In other words, the LSA between $R^2_n$ and $R^2_c$ is guaranteed
by those between $R^2_n$ and $Q^4_c$, and between $R^2_n$ and $Q^4_{cn}$
whose relationships are clear, as in Eq.(\ref{4thm}).

\begin{figure}[ht]
\centering{%
\includegraphics[bb=0 0 216 162]{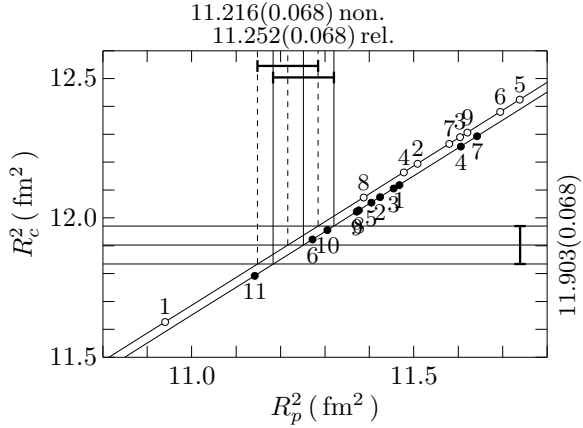}
}
\caption{The mean square radius(msr) of the charge density($R^2_c$) as a function of
the msr of the point proton density($R^2_p$) in $^{40}$Ca.
The closed(open) circles are calculated in various relativistic(non-relativistic)
mean field models. The number assigned to each circle represents a model use in
the calculations as explained in the text. The lines of the least square fitting
are shown for the relativistic and non-relativistic models separately. Those lines
cross the horizontal lines which express the experimental value with error indicated
on the left-hand side of the figure. Their intersection points are described on the top
of the figure for the relativistic and non-relativistic models, respectively.
}
\label{RpRc-Ca40}
\end{figure}

First, let us analyze the msr of the point proton distribution.
Figure \ref{RpRc-Ca40} is a typical example of the LSA in the present paper.
It shows $R_c^2(R^2_{c,{\rm nr}})$ as a function of $R_p^2(R^2_{p,{\rm nr}})$
calculated for $^{40}$Ca, 
by 11 relativistic models indicated with the filled circles
and by 9 non-relativistic ones with the open circles.
In the horizontal axis,  $R_p^2$ should be read as $R^2_{p,{\rm nr}}$,
while in the vertical one, $R_c^2$ as $R^2_{c,{\rm nr}}$ for the
non-relativistic models.
In all the following figures, the horizontal and  vertical axes
should be read in the same way for the non-relativistic models.
For other cases also, the same notations will be used often
for the relativistic and non-relativistic models without notice,
when the meaning of the notation is clear.

Each circle in the figure has the number to specify the corresponding nuclear model.
The numbers 1 to 11 of the filled circles represent the relativistic nuclear models,
as 1 L2\cite{sw1}, 2 NLB\cite{sw1}, 3 NLC\cite{sw1}, 4 NL1\cite{nl1}, 
5 NL3\cite{nl3}, 6 NL-SH\cite{nlsh}, 7 NL-Z\cite{nlz}, 8 NL-S\cite{nls},
9 NL3II\cite{nl32}, 10 TM1\cite{tm1} and 11 FSU\cite{fsu},
while those from 1 to 9 of the open circles stand for the non-relativistic nuclear models,
as 1 SKI\cite{sk1}, 2 SKII\cite{sk1},  3 SKIII\cite{sk3}, 4 SKIV\cite{sk3},
5 SkM$^\ast$\cite{skm}, 6 SLy4\cite{sly4}, 7 T6\cite{st6}, 8 SGII\cite{sg2},
and 9 Ska\cite{ska}. 
These designations of the circles and the numbers will be used throughout
the present paper.

For the relativistic models, the values of the nucleon mass are taken
from their references, 
while for the non-relativistic models, $M=$939 MeV is used.
These choices are not essential for the following discussions.

Figure \ref{RpRc-Ca40} shows the two LSL's calculated using the phenomenological models.
The one is for the relativistic framework, and the other is for the non-relativistic one.
If model-frameworks are different from each other, thus
their LSL's are not the same usually.
In the non-relativistic models, all the calculated values should be on the line,
since $R^2_{c,\,{\rm nr}}$ is proportional to $R^2_{p,{\rm nr}}$ and others are
constant in the mean field models for $^{40}$Ca, as seen in Eq.(\ref{correct}).
The equation of the line is given by $R^2_{c,{\rm nr}} = 1.0000R^2_{p,{\rm nr}}+0.6863$
with $\sigma=0.0000$ fm$^2$.
Here, $\sigma$ denotes the standard deviation,
which is defined as
\[
n\sigma^2=\sum_{i=1}^n(y_i-ax_i-b)^2,
\]
where $n$ represents the number of the samples, $y_i$ and $x_i$ the calculated values
like those of $R_c^2$ and $R_p^2$, and $ax_i+b$ is given by the equation of the LSL, $y=ax+b$.
The line of the relativistic models is given by
$R_c^2 = 1.0003R_p^2+0.6472$ with $\sigma=0.0002$ fm$^2$.
The coefficient of $R^2_p$ is a little different from 1 and the value of $\sigma$ is not
exactly 0, because of the contribution from the spin-orbit density
in the relativistic models in Eq.(\ref{rmsr}).

The experimental value of $R_c^2$ with the error is indicated on the
right-hand side, 11.903(0.068) fm$^2$, corresponding to $R_c$=3.450(0.010) fm\cite{emrich}.
The intersection points of the LSL's with the line for the experimental value
are shown on the top as $R^2_p=$11.252(0.068) fm$^2$
and $R^2_{p,{\rm nr}}$=11.216(0.068) fm$^2$,
corresponding to $R_p=3.354(0.011)$ fm and $R_{p,{\rm nr}}=3.349(0.010)$ fm,
in relativistic and non-relativistic models, respectively.
They are considered to be the accepted values of $R_p$ and $R_{p,{\rm nr}}$
from the LSA between  $R_p^2(R^2_{p,{\rm nr}})$
and $R_c^2(R^2_{c,{\rm nr}})$ in the mean field approach.

Since even in the relativistic models, the contribution from the spin-orbit density is
small in $^{40}$Ca, the value of the intercept, 0.6472 of the relativistic LSL, is almost equal to
the value of $r^2_p+(r_+^2-r_-^2)=0.6531$ fm$^2$ according to Eq.(\ref{rmsr}).
In the non-relativistic models, the additional contribution to the
intercept may come from  $3/(4M^2)=0.0331$ fm$^2$, as the sum of them,
0.6531+0.0331= 0.6862, is equal to the value of the intercept, 0.6863 of the LSL,
except for the numerical error of the last digit.
In fact, the value of $3/(4M^2)$ is added to that of $R^2_{p, {\rm nr}}$
as a relativistic correction in Eq.(\ref{r}) by hand, since
it has not been taken into account in some previous papers\cite{sk1, sk3,st6}. 
If the correction is considered to be already included implicitly
in the non-relativistic interaction parameters which are fixed by
experimental values, the difference between the lines of the two frameworks
would almost disappear.
This interpretation of the intercept in the non-relativistic models, however, 
may be a part of the solution. In the relativistic models, the term corresponding to
$3/(4M^2)$ is contained as $3/(4M^{\ast 2})$, $M^\ast$ being $\sim 0.6M$,
as shown in the second term of Eq.(\ref{rr}).  
Additionally, in that case, one must accept $R^2_p=R^2_{p, {\rm nr}}$, which seems not
to be reasonable at this stage.

Figure \ref{RpRc-Ca40} shows that 6(NL-SH)\cite{nlsh} and 10(TM1)\cite{tm1}
well reproduce the experimental value within the error of $R_c^2$, but
it is not necessary for them to explain other experimental values.
For example, the former yields the nuclear matter incompressibility as
$K=355$ MeV. It fails to describe  the isoscalar giant monopole resonance states
which require $K\approx 230$ MeV\cite{nl3,nlsh}.
The latter predicts the value of $R_p^2$ for $^{208}$Pb
which overestimates the experimental one, as seen later.
Thus, there is no reason to choose NL-SH or TM1 as the best
among the phenomenological models.

Moreover, it should be noticed in Figure \ref{RpRc-Ca40} that all the values
of $R^2_{c,{\rm nr}}$ evaluated with the non-relativistic models do not agree
with experiment, and are not on the band of 11.903(0.068) obtained
from the Fourier-Bessel analysis of electron scattering data\cite{emrich}.
If one were to compare the calculated values in the relativistic models with
those in the non-relativistic models, one would conclude that
the average value of $R^2_p$ is smaller than that of $R^2_{p, {\rm nr}}$.

In the following, all figures in the present paper will be shown in a similar
way as Figure  \ref{RpRc-Ca40}. The equations of the LSL's will be
listed in the table at the end of each subsection together with the values of
$\sigma$, for convenience to compare them with one another.
If the value of $\sigma$ is large enough to depict the $\pm \sigma$
area, then it will be shown in the figures explicitly.
In Figure \ref{RpRc-Ca40}, its area is not seen, since the value of $\sigma$ is too small.
The accepted regions for the values of the moments
will be indicated on the top of each figure as in Figure \ref{RpRc-Ca40}.
The spread of the region taking into account $\sigma$ is described
in the parenthesis following the mean accepted value.
In the present section, however, the accepted regions will be discussed neglecting
$\sigma$, in order to make easier the comparison with the previous
discussions without $\sigma$ in Refs.\cite{abra,roca} and in order to focus the present
discussions mainly on the LSL's themselves.
The accepted regions which take account of $\sigma$
will be summarized in the last section separately, and will be discussed in Appendix.

\begin{figure}[h]
\begin{minipage}{7.7cm}
\includegraphics[bb=0 0 211 161]{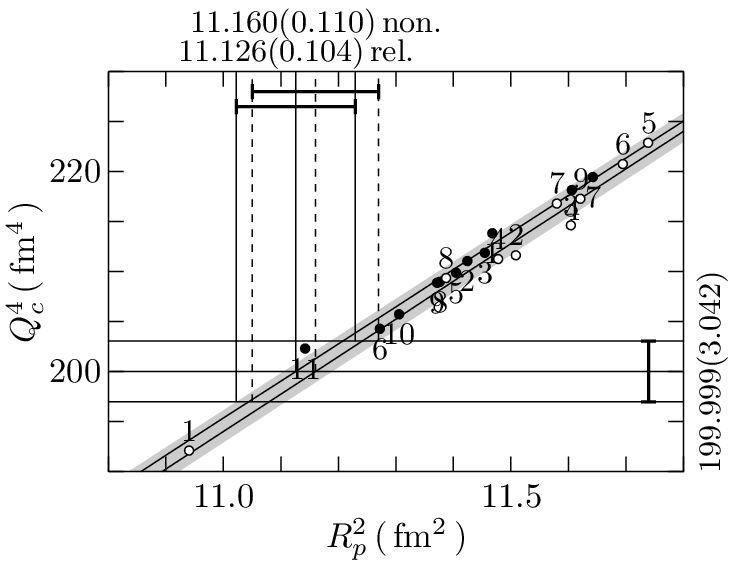}%
 \caption{The same as Figure \ref{RpRc-Ca40}, but for the fourth-order moment($Q^4_c$) of the
 nuclear charge density against $R^2_p$ in $^{40}$Ca. The gray area denotes the standard deviation
 of the calculated values from the least square lines.}
\label{RpQc-Ca40}
\end{minipage}\hspace{0.3cm}%
\begin{minipage}{7.7cm}
\includegraphics[bb=0 0 211 162]{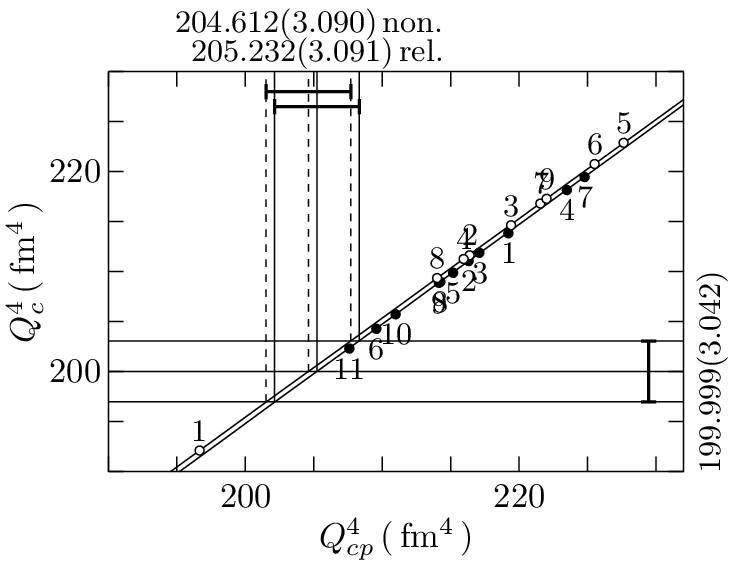}
\caption{The same as Figure \ref{RpRc-Ca40}, but for $Q^4_c$ against the fourth-order
moment of the proton charge density ($Q^4_{cp}$) in $^{40}$Ca.
}
\label{QcpQc-Ca40}
\end{minipage}
\end{figure}

Figure \ref{RpQc-Ca40} shows the LSL's between $R_p^2$ and $Q_c^4$.
The relativistic line is given by $Q_c^4=37.1394R_p^2-213.2060$,
and the non-relativistic one by $Q_c^4=37.5401R_p^2-218.9541$.
The grey shaded area denotes the $\pm \sigma$ spread, although the relativistic and
non-relativistic ones are overlapped in the present case.
The two lines cross the experimental region at 11.126(0.082) and 11.160(0.081) fm$^2$,
respectively. As noted in the above, the numbers of the parentheses are different from
those on the top of the figure, depending on whether or not $\sigma$ is taken into
account. It will be explained in Appendix how the errors are increased owing to $\sigma$

\begin{figure}[ht]
\centering{\includegraphics[bb=0 0 219 161]{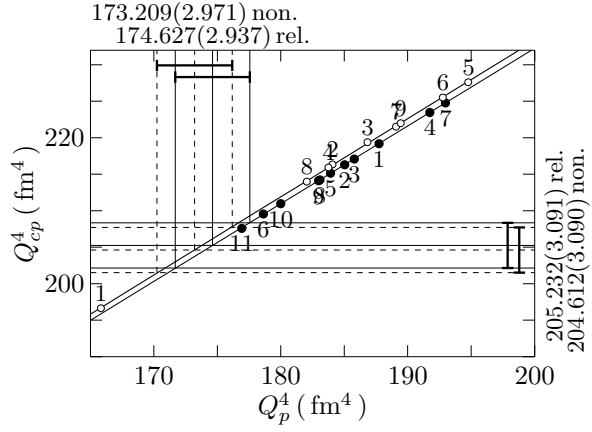}}
\caption{The same as Figure \ref{RpRc-Ca40}, but for $Q^4_{cp}$ against the fourth-order
moment of the point proton density ($Q^4_p$) in $^{40}$Ca.
}
\label{QpQcp-Ca40}
\end{figure}

Figure \ref{QcpQc-Ca40} shows the LSL's for $Q^4_{cp}$ and $Q^4_c$.
Both relativistic and non-relativistic models yield the well
defined straight lines. The values of $\sigma$ is too small
to show the $\pm\sigma$ area.
They provide the value of $Q^4_{cp}$ to be 205.232(3.055) and 204.612(3.064) fm$^4$
for relativistic and non-relativistic frameworks, respectively.
The difference between these values is mainly due to $Q^4_p$ and $Q_{4W_p}$ in the two models.
On the one hand, Table 2 shows that $Q_{4W_p}$ in the non-relativistic model
is negligible, but not in the relativistic models.
On the other hand, the value of $Q^4_p$ by SLy4(6) is larger than those of
NL3(5) and NL-SH(6) in Table 2, but the LSA between $Q^4_p$ and
$Q^4_{cp}$ in Figure \ref{QpQcp-Ca40} shows that the accepted value of $Q^4_p$ is
174.627(2.868) fm$^4$ in the relativistic framework, while that of $Q^4_{p, {\rm nr}}$
173.209(2.860) fm$^4$ in the non-relativistic one. All values in the non-relativistic models  
are outside of the accepted region and the one of $Q^4_{p, {\rm nr}}$ by SLy4(6)
is large, compared with others.

As seen in Figure \ref{RpRc-Ca40}, the value of $R_p^2$ of FSU(11)
underestimates the experimental values.
In Figure \ref{RpQc-Ca40} and \ref{QpQcp-Ca40}, however, the only FSU yields the values
of $Q_{cp}^4$ and $Q_p^4$ within the error of the experimental values.
FSU(11) has two additional parameters, compared with other
relativistic models\cite{fsu}.
Thus, by employing the experimental value of $Q^4_c$, the LSA
makes it possible to explore not only $R_p^2$, but also $Q_p^4$
which provides more information on the nuclear surface.

\begin{figure}[ht]
\centering{\includegraphics[bb=0 0 219 162]{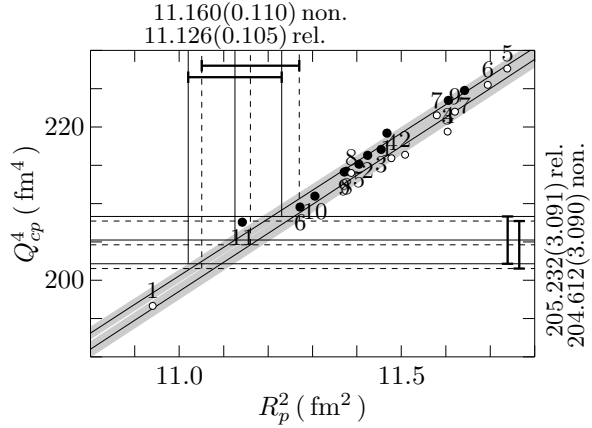}}
\caption{The same as Figure \ref{RpRc-Ca40}, but for $Q^4_{cp}$ against $R^2_p$ in $^{40}$Ca.
The gray area denotes the standard deviation
 of the calculated values from the least square lines.
}
\label{RpQcp-Ca40}
\end{figure}

Using the above accepted values of $Q^4_{cp}$ in Figure \ref{QcpQc-Ca40},
the LSL's are obtained for $R^2_p$  as in Figure \ref{RpQcp-Ca40}.
The equations of the lines and the value of $\sigma$ are a little different from
those in Figure \ref{RpQc-Ca40}, as listed in Table 4,
but the accepted values of $R^2_p$ are almost the same as those in Figure \ref{RpQc-Ca40},
for both relativistic and non-relativistic models.
The relationship between Figure \ref{RpQc-Ca40} and Figure \ref{RpQcp-Ca40} will be discussed
in more detail in Appendix.

Table 4 shows that the values of the slopes of the LSL's
in Figure \ref{RpRc-Ca40}, \ref{QcpQc-Ca40} and \ref{QpQcp-Ca40} are 
almost equal to 1, as expected from Eq.(\ref{4thm}) and (\ref{n4thm}),
but in Figure \ref{RpQc-Ca40} and \ref{RpQcp-Ca40},
they are far from the value of the coefficient, $(10/3)r^2_p=2.564$ fm$^{2}$,  of $R^2_p$
in those equations. This result is owing to the fact that
when $R^2_p$ increases, the main component $Q^4_p$ in $Q^4_c$ and $Q^4_{cp}$ also increases
together, satisfying at least the identity of the variance, $Q^4_p-R^4_p > 0$.
Indeed, the LSL's of Figure \ref{QpQcp-Ca40} and \ref{RpQcp-Ca40} provide
the relationships,
\begin{align}
&Q^4_p\,\,\,=34.9833R^2_p-214.5822, \quad \nonumber\\
&Q^4_{p,{\rm nr}}=35.3037R^2_{p,{\rm nr}}-220.7904\,. \quad \label{qprp}
\end{align}
Thus, the values of the slopes
in Figure \ref{RpQc-Ca40} and \ref{RpQcp-Ca40} are nearly equal to the values of the
above equations. It is noticeable that $R^2_p $ changes independently of other elements
in Eq.(\ref{rmsr}) for Figure \ref{RpRc-Ca40}, while it varies together with other
components in Eq.(\ref{4thm}) for Figure \ref{RpQc-Ca40} and \ref{RpQcp-Ca40}.
Nevertheless, the LSA provides the accepted ranges of $R^2_p$
which are consistent with one another in Figure \ref{RpRc-Ca40},
\ref{RpQc-Ca40} and \ref{RpQcp-Ca40}.

Finally, from Figures \ref{RpRc-Ca40}, \ref{RpQc-Ca40} and \ref{RpQcp-Ca40}, on the one hand,
the common accepted region of $R^2_p$ in the relativistic framework is decided to be
$\mathcal{R}_p=11.184 \sim 11.208$ fm$^2$, which corresponds to $R_p= 3.344 \sim 3.348$ fm.
The lower bound is obtained from Figure \ref{RpRc-Ca40}
and the upper bound from Figure \ref{RpQc-Ca40} and \ref{RpQcp-Ca40}.
On the other hand, for the non-relativistic models, they are obtained to be 
$\mathcal{R}_{p, {\rm nr}}=11.148 \sim 11.241$ fm$^2$, yielding $R_{p,{\rm nr}} =3.339 \sim 3.353$ fm.
The lower and the upper bound are from Figure \ref{RpRc-Ca40}, and \ref{RpQc-Ca40}
and \ref{RpQcp-Ca40}, respectively.
Thus, it makes the accepted region of $R^2_p$ narrower
to take into account the three LSL's together. 

\begin{figure}[ht]
\centering{\includegraphics[bb=0 0 216 161]{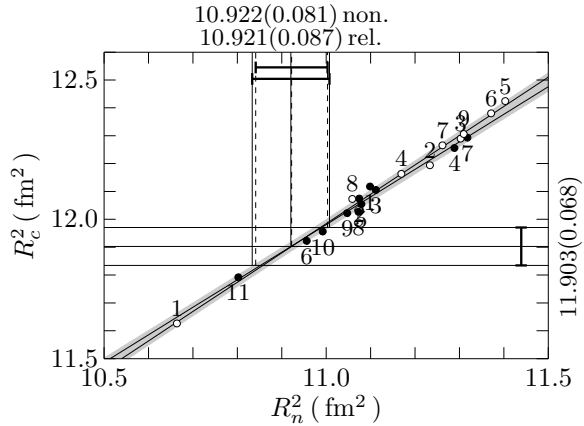}}
\caption{The same as Figure \ref{RpRc-Ca40}, but for $R^2_c$ against the msr of the
 point neutron density($R^2_n$) in $^{40}$Ca. The gray area denotes the standard deviation
 of the calculated values from the least square lines.}
\label{RnRc-Ca40}
\end{figure}

Next, we analyze the msr of the neutron distribution in $^{40}$Ca
in the same way as for $R^2_p$.
Figure \ref{RnRc-Ca40} shows the relationship between $R_n^2$ and $R_c^2$.
The LSL for the relativistic models is given by 
$R_c^2=0.9897R_n^2+1.0942$,
and for the non-relativistic ones by $R_c^2=1.0527R_n^2+0.4046$.
The two lines are separated, but the gray areas are overlapped
with each other in spite of their small values of $\sigma$.
The values of their slopes are almost equal to 1, because of $R^2_n\approx R^2_p$,
but the meaning of the intercepts is not clear, unlike that for $R^2_p$.
The LSL's yield almost the same value of $R^2_n$ for the relativistic
and non-relativistic models, as 10.921(0.069) and 10.922(0.065)  fm$^2$, respectively.

These lines reflect the fact that $R_n^2$ strongly correlates with
and increases with $R^2_p$ in Figure \ref{RpRc-Ca40}.
Unlike the case of $R^2_p$, however, $R^2_c$ is not described explicitly in term of
$R^2_n$, so that it is not trivial whether or not all the calculated ones of
$R^2_c$ are on the LSL's as a function of $R^2_n$.
The present method provides us with the accepted values of $R^2_n$ within the narrow ranges
in the relativistic and non-relativistic frameworks,
even though most of their calculated values of $R^2_c$ do not reproduce the experimental one exactly.
Thus, the LSA provides us with common constraints on the
mean field models which are almost independent of their parameterizations.

Equations of the LSL's in Figure \ref{RpRc-Ca40} and \ref{RnRc-Ca40} provide the relationship,
\begin{align}
&R^2_p\,\,\,\,=0.9894R^2_n+0.4469, \quad \nonumber\\
&R^2_{p,{\rm nr}}=1.0527R^2_{n,{\rm nr}}-0.2817\,.\label{rprn}
\end{align}
In using the values of $R^2_p(R^2_{p, {\rm nr}})$ determined
in Figure \ref{RpRc-Ca40} as the pseudo experimental ones,
the LSA between $R^2_n(R^2_{p, {\rm nr}})$ and $R^2_p(R^2_{p,{\rm nr}})$ provides
the same values of $R^2_n$ and $R^2_{n,{\rm nr}}$ as those from Figure \ref{RnRc-Ca40},
as expected, and their LSL's to be given by 
\begin{align}
&R^2_p\,\,\,\,=0.9893R^2_n+0.4480, \quad \nonumber\\
&R^2_{p,{\rm nr}}=1.0527R^2_{n,{\rm nr}}-0.2816\,,\label{rprn1}
\end{align}
with $\sigma=0.0180$ and 0.0167 fm$^2$, respectively.
Eq.(\ref{rprn1}) is almost the same as Eq.(\ref{rprn}).
Eq.(\ref{rprn}) and (\ref{rprn1}) together with Eq.(\ref{qprp}) are additional
results obtained from the present LSA.

\begin{figure}[ht]
\centering{\includegraphics[bb=0 0 211 162]{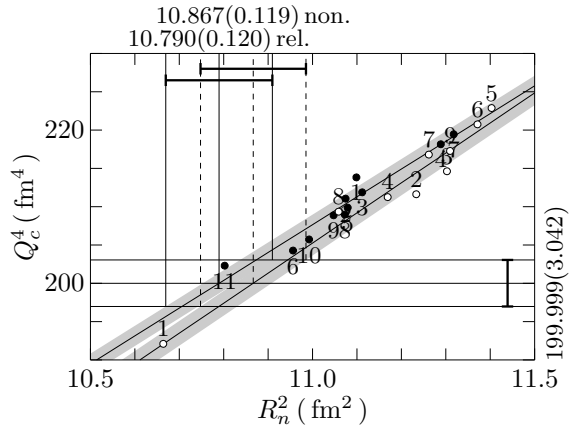}}
\caption{The same as Figure \ref{RpRc-Ca40}, but for $Q^4_c$ against $R^2_n$
in $^{40}$Ca. The gray area denotes the standard deviation
 of the calculated values from the least square lines.
}
\label{RnQc-Ca40}
\end{figure}

Figure \ref{RnQc-Ca40} shows the relationship between $R_n^2$ and $Q_c^4$.
The analysis between these quantities is a typical example
which is performed in the same way as for  $R_n$ and $A_{\rm pv}$
in the parity-violating electron scattering\cite{abra}.
In the present case, however, the structure of $Q^4_c$ is well defined, as 
in Eq.(\ref{4thm}) and (\ref{n4thm}), and the meaning of each contributed
component is apparent.
Moreover, among the neutron moments, the only $R^2_n$ contributes to $Q^4_c$.
In combining with the analysis on $R^2_p$, the value of $\delta R =R_n-R_p$
is obtained on the same basis, as will be seen later,
although in the parity-violating electron scattering also,
the analysis on $R_p$ and $A_{\rm pv}$ would be possible. 

The two LSL's in Figure \ref{RnQc-Ca40} are described by by $Q_c^4=36.2764R_n^2-191.4118$,
and $Q_c^4=39.2239R_n^2-226.2309$
for the relativistic and non-relativistic frameworks, respectively. 
According to Eq.(\ref{4thm}), $R_n^2$ in $Q_{2n}$  contributes to
$Q_c^4$ with a negative coefficient, but the LSL's have
positive ones. This implies that not only $R_p^2$, but also $Q^4_{p}$
in $Q^4_c$ increases with $R_n^2$ in these model calculations, 
Indeed, their slopes are similar to those in Figure \ref{RpQc-Ca40} and \ref{RpQcp-Ca40},
owing to Eq.(\ref{qprp}) and(\ref{rprn}).
The value of $\sigma$ in Figure \ref{RnQc-Ca40} is the largest in $^{40}$Ca,
but most of the calculated values of $Q^4_c$ are within the $\pm \sigma$ areas.
The accepted region of $R_n^2$ is obtained to be 10.790(0.084) fm$^2$ for the relativistic models,
and 10.867(0.078) fm$^2$ for the non-relativistic models, in neglecting $\sigma$.

The LSL's of Figure \ref{RpQc-Ca40} and  \ref{RnQc-Ca40} 
provide the relationship as
\begin{align}
&R^2_p\,\,\,\,=0.9768R^2_n+0.5868, \quad \nonumber\\
&R^2_{p,{\rm nr}}=1.0449R^2_{n,{\rm nr}}-0.1938\,.\label{rprn2}
\end{align}
These are slightly different from Eq.(\ref{rprn}) obtained from Figure \ref{RpRc-Ca40}
and \ref{RnRc-Ca40}, but the difference is within the experimental errors.
It should be noted that Eq.(\ref{rprn2}) is derived using a small contribution of the term
with $R^2_n$ to $Q^4_c$ in Eq.(\ref{4thm}) and (\ref{n4thm}). Its contribution is less than
$5\%$, but the change of the $R^2_n$-value induces the change of the contribution from
other components to $Q^4_c$.

According to the relationship, $Q^4_{cn}=Q^4_{cp}-Q^4_c$,
Figure \ref{QcpQc-Ca40} provides the accepted values
of $Q^4_{cn}$ to be 5.233(0.013) and 4.613(0.022) fm$^4$ for the relativistic
and non-relativistic frameworks, respectively,
but neglecting $\sigma$, as mentioned before.
If the standard deviation is taken into account, they are given
by 5.233(0.049) and 4.613(0.047) fm$^4$.
The LSA requires Eq.(\ref{4thm}) to reproduce
the experimental value of $Q^4_c$ in the relativistic models,
while Eq.(\ref{n4thm}) to explain the same value in the non-relativistic models.
Hence, the components of those two equations should satisfy
\begin{equation}
Q^4_{cn}-Q^4_{cn,{\rm nr}}=Q^4_{cp}-Q^4_{cp,{\rm nr}}.\label{qrqn}
\end{equation}
On the one hand,
the difference between the above two values, 5.233-4.613 =0.620, is for the left-hand side
of Eq.(\ref{qrqn}), and stems from the different values of the contributions
from ($Q_{2n}+Q_{2W_n}+Q_{4W_n}$) to $Q^4_{cn}$ and from the corresponding terms
to $Q^4_{cn,{\rm nr}}$.
On the other hand, the value of the difference between $Q^4_{cp}$
and  $Q^4_{cp,{\rm nr}}$ 
in the right-hand side of Eq.(\ref{qrqn}) is obtained 
from Figure \ref{QcpQc-Ca40} to be, of course, 205.232-204.612=0.620,
but it stems mainly from the difference between
$Q^4_p$ and $Q_{4W_p}$ in $Q^4_{cp}$ and those in $Q^4_{cp,{\rm nr}}$.
Thus, it is required in the present analysis
that these proton contributions in the right-hand side of Eq.(\ref{qrqn})
are exactly equal to the neutron
ones in the left-hand side from the different origin,
in order for both frameworks to reproduce the same experimental value.

\begin{figure}[h]
\begin{minipage}{7.7cm}
\includegraphics[bb=0 0 211 161]{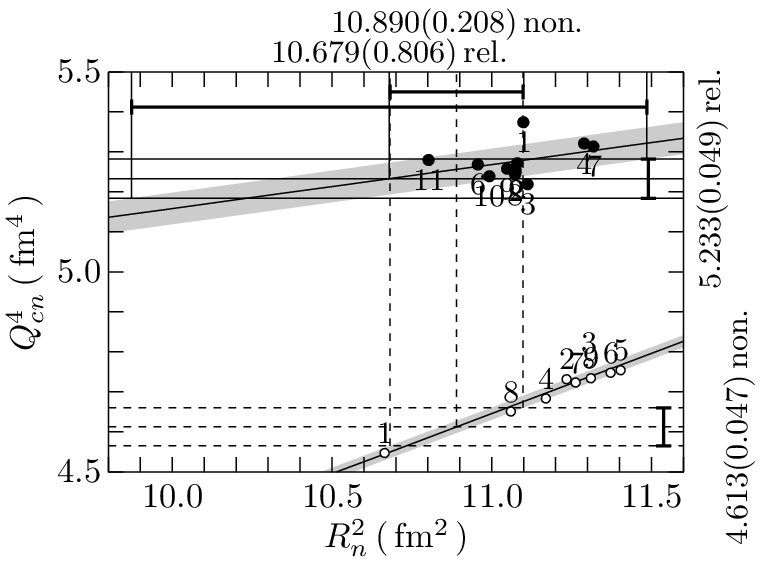}
\caption{The same as Figure \ref{RpRc-Ca40}, but for the fourth-order moment of
 the neutron charge density($Q^4_{cn}$) against $R^2_n$
in $^{40}$Ca. The gray area denotes the standard deviation
of the calculated values from the least square lines.
}
\vspace{0.4cm}
\label{RnQcn-Ca40}
\end{minipage}\hspace{0.3cm}%
\begin{minipage}{7.8cm}
\includegraphics[bb=0 0 219 161]{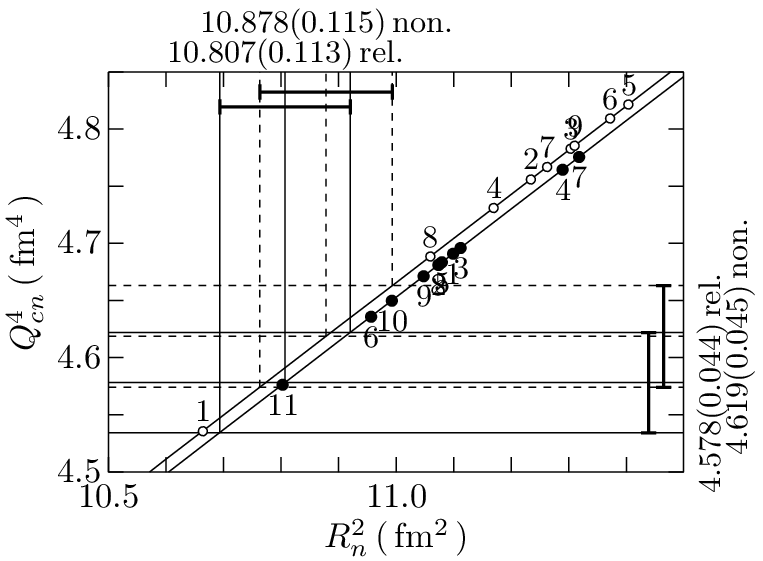}
\caption{The same as Figure \ref{RpRc-Ca40}, but for $Q_{cn}^4$ calculated without the 
spin-orbit density against $R^2_n$ in $^{40}$Ca. For details, see the text.
}
\label{RnQcn-Ca40-noLS}
\end{minipage}
\end{figure}

Using the above values for $Q^4_{cn}$ without $\sigma$, Figure \ref{RnQcn-Ca40} determines
the accepted regions of $R_n^2$ by the LSL's between $R^2_n$
and  $Q^4_{cn}$. The relativistic line is given by $Q_{cn}^4=0.1100R_n^2+4.0586$,
giving $R_n^2$ to be 10.679(0.116)fm$^2$, 
and the non-relativistic one
by $Q_{cn}^4=0.3010R_n^2+1.3345$,
yielding $R_{n,{\rm nr}}^2$ to be 10.890(0.072) fm$^2$.
The calculated values of $Q^4_{cn,{\rm nr}}$ in non-relativistic models are well on the line,
while the relativistic ones are distributed around the line.
The reason of this fact is understood as follows.
As seen in Table 2, the value of $Q^4_{cn}$ is dominated
by $Q_{2n}$ in both relativistic and non-relativistic models,
but the relativistic models predict the non-negligible contribution
from $Q_{4W_n}$. The small value of the slope, 0.1100, also shows this fact,
in comparing with the value, 0.3010, in the non-relativistic models.

The effects of the spin-orbit density are more clearly seen in Figure \ref{RnQcn-Ca40-noLS}
which is obtained by neglecting them in Figure \ref{RnQcn-Ca40}.
The LSL's in this case are given by
$Q_{cn}^4=0.3867R_n^2+0.3996$ for the relativistic models,
and $Q_{cn}^4=0.3867R_n^2+0.4124$ for the non-relativistic ones.
According to Eq.(\ref{4thm}) and (\ref{n4thm}), their slopes of the
lines should be $-(10/3)(r^2_+-r^2_-)N/Z=0.3867$.
The intercepts are in the relativistic models,
\[
 -\,\frac{5}{2}(r^4_+-r^4_-)\frac{N}{Z}=0.3996,
\]
while in the non-relativistic models,
\[
-\,\frac{5}{2}(r^4_+-r^4_-)\frac{N}{Z}
-\frac{10}{3}(r^2_+-r^2_-)\frac{N}{Z}\frac{3}{4M^2}=0.4124.
\]
The additional term of the above equation in the non-relativistic models stems from
\[
 Q_{2n,{\rm nr}}=-\,\frac{10}{3}(r^2_+-r^2_-)\frac{N}{Z}
\left(
\avr{r^2}_{n,{\rm nr}}+\frac{3}{4M^2}
\right),
\]
which is obtained by  Eq.(\ref{equivalent}) in neglecting the spin-orbit density.
The lines in Figure \ref{RnQcn-Ca40-noLS} are similar to each other,
but yield different values of $R_n^2$ for the relativistic and the non-relativistic models,
since, in addition to the different values of the intercepts,
the values of $Q^4_{cn}$ for the two frameworks are different from each other,
as indicated on the right-hand side of Figure \ref{RnQcn-Ca40-noLS}.
Thus, in comparing Figure \ref{RnQcn-Ca40} with \ref{RnQcn-Ca40-noLS},
the role of the spin-orbit density in $Q^4_c$ may be understood at a glance.

For the relativistic models, unfortunately, Figure \ref{RnRc-Ca40}, \ref{RnQc-Ca40}
and \ref{RnQcn-Ca40} have no common accepted region of $R_n^2$, although two of them do.
The accepted region of Figure \ref{RnQc-Ca40} from 10.706 to 10.874 fm$^2$ contains 
a part of that in Figure \ref{RnRc-Ca40}, 10.852 to 10.990 fm$^2$,
and in Figure \ref{RnQcn-Ca40}, 10.563 to 10.795 fm$^2$.
This is the only exception where there is no common region in the present analysis.
In fact, when $\sigma$ is taken into account, the accepted region of $R^2_n$
given by the relationship between $R^2_n$ and $Q^4_{cn}$ contains other two regions
provided by those between $R^2_n$ and $R^2_c$ and between $R^2_n$ and $Q^4_c$.
Invoking this fact, Figure \ref{RnRc-Ca40} and \ref{RnQc-Ca40} are employed to
determine the accepted region in the present case.
Then, the accepted range of $R^2_n$ is given by $\mathcal{R}_n=10.852\sim 10.874$ fm$^2$,
yielding $R_n= 3.294\sim 3.298$ fm.

\begin{figure}[ht]
\centering{\includegraphics[bb=0 0 211 162]{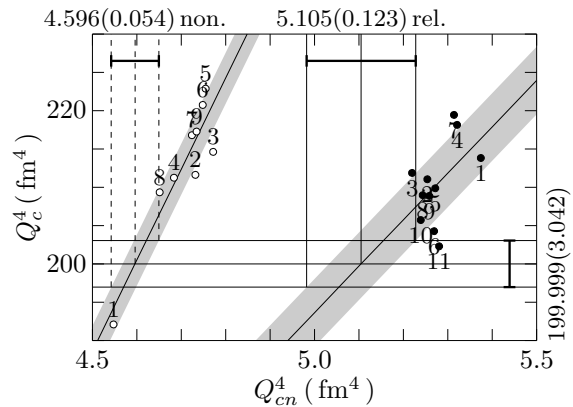}}
\caption{The same as Figure \ref{RpRc-Ca40}, but for $Q^4_c$ against $Q_{cn}^4$
in $^{40}$Ca. The gray area denotes the standard deviation
 of the calculated values from the least square lines.
}
\label{QcnQc-Ca40}
\end{figure}

\begin{table}
\renewcommand{\arraystretch}{1.3}
\begin{tabular}{|c|c|c||r|r|r||r|r|r|} \hline
\multicolumn{3}{|c||}{$^{40}$Ca}  & \multicolumn{3}{c||}{Rel.} &
\multicolumn{3}{c|}{Non.} \\ \hline
Fig.  & $y$     & $x$     & \multicolumn{1}{c|}{$a$} & \multicolumn{1}{c|}{$b$} &
\multicolumn{1}{c||}{$\sigma$} & \multicolumn{1}{c|}{$a$} & \multicolumn{1}{c|}{$b$} &
\multicolumn{1}{c|}{$\sigma$} \\ \hline
\ref{RpRc-Ca40} &  $R_c^2$ & $R_p^2$ &
$ 1.0003$ & $    0.6472$ & $ 0.0002$ & $ 1.0000$ & $    0.6863$ & $ 0.0000$ \\ \hline
\ref{RpQc-Ca40} & $Q_c^4$ & $R_p^2$ & 
$37.1394$ & $ -213.2060$ & $ 0.8028$ & $37.5401$ & $ -218.9541$ & $ 1.0780$ \\ \hline
\ref{QcpQc-Ca40}& $Q_c^4$ &  $Q_{cp}^4$ &
$ 0.9958$ & $   -4.3744$ & $ 0.0363$ & $ 0.9929$ & $   -3.1634$ & $ 0.0256$ \\ \hline
\ref{QpQcp-Ca40}& $Q_{cp}^4$ &  $Q_{p}^4$ &
$ 1.0653$ & $   19.2025$ & $ 0.0373$ & $ 1.0713$ & $   19.0527$ & $ 0.0934$ \\ \hline
\ref{RpQcp-Ca40}&  $Q_{cp}^4$ &  $R_{p}^2$ &
$37.2677$ & $ -209.3919$ & $ 0.8286$ & $37.8209$ & $ -217.4801$ & $ 1.0623$ \\ \hline
\ref{RnRc-Ca40}&  $R_{c}^2$ &  $R_{n}^2$ &
$ 0.9897$ & $    1.0942$ & $ 0.0179$ & $ 1.0527$ & $    0.4046$ & $ 0.0167$ \\ \hline
\ref{RnQc-Ca40}&  $Q_{c}^4$ &  $R_{n}^2$ &
$36.2764$ & $ -191.4118$ & $ 1.3097$ & $39.2239$ & $ -226.2309$ & $ 1.6123$ \\ \hline
\ref{RnQcn-Ca40}&  $Q_{cn}^4$ &  $R_{n}^2$ &
$ 0.1100$ & $    4.0586$ & $ 0.0394$ & $ 0.3010$ & $    1.3345$ & $ 0.0152$ \\ \hline
\ref{RnQcn-Ca40-noLS}&  $Q_{cn}^4$ &  $R_{n}^2$ &
$ 0.3867$ & $    0.3996$ & $ 0.0000$ & $ 0.3867$ & $    0.4124$ & $ 0.0000$ \\ \hline
\ref{QcnQc-Ca40}&  $Q_{c}^4$ & $Q_{cn}^4$ &
 $60.6119$ & $ -109.4354$ & $ 4.4084$ & $118.8061$ & $ -346.0516$ & $ 3.3254$ \\ \hline 
\end{tabular} 
\caption{The least square line $y(x)=ax+b$ and the standard deviation $\sigma$ depicted
 in Figure \ref{RpRc-Ca40} to \ref{RnQcn-Ca40-noLS} for the relativistic(Rel.)
 and the non-relativistic(Non.) models.}
\end{table}

For non-relativistic models, Figure \ref{RnRc-Ca40}, \ref{RnQc-Ca40}
and \ref{RnQcn-Ca40} provide the common accepted region
of $R_{n,{\rm nr}}^2$ with the lower bound from Figure \ref{RnRc-Ca40}
and the upper bound from Figure \ref{RnQc-Ca40} as
$\mathcal{R}_{n, {\rm nr}}= 10.857\sim 10.945$ fm$^2$,
yielding $R_{n,{\rm nr}}=3.295$ to 3.308 fm.
The accepted region from 10.818 to 10.962 fm$^2$ in Figure \ref{RnQcn-Ca40}
contains the above common one.
Thus, it seems reasonable to take into account the three kinds of the LSL's for discussions of $R_n$. 
In the present case, Figure \ref{RnRc-Ca40} makes the accepted region narrower.

According to the above analysis, $R_p$ is predicted to be larger a little
than $R_n$ in both the relativistic and non-relativistic models, as
expected from the Coulomb energy contribution to the total energy of the nucleus. 
The skin thickness defined by $\delta R=R_n-R_p$ is given to be $-(0.046\sim 0.054)$ fm
in the relativistic framework, and $-(0.031\sim 0.058)$ fm in the non-relativistic framework.
Eq.(\ref{rprn}),(\ref{rprn1}) and (\ref{rprn2}) are consistent with these values.

Before closing this subsection, it should be mentioned why the values of $Q^4_{cn}$ used
in Figure \ref{RnQcn-Ca40} have been derived from the relationship $Q^4_{cn}=Q^4_{cp}-Q^4_c$,
but not from the LSA between $Q^4_{cn}$ and $Q^4_c$.
The reason is as follows. On the one hand, the values of $Q^4_{cp}$ are well determined by
the experimental values of $Q^4_c$, according to Figure \ref{QcpQc-Ca40}.
As shown in Table 4, the LSL's in Figure \ref{QcpQc-Ca40} are described
with small values of $\sigma$. On the other hand, in the case of the relationship
between $Q^4_{cn}$ and $Q^4_c$, the LSA seems not to be useful,
in particular, for the relativistic models. Figure \ref{QcnQc-Ca40} shows this fact in $^{40}$Ca
as an example. Most of the closed circles are concentrated in the same region
around $Q^4_{cn}$=5.25 fm$^4$, and the LSL is dominated by a few rest of the models.
Such a distribution of the circles is not appropriate for the analysis by the LSA.
Indeed, the value of $\sigma$ of the relativistic line is large as 4.4084 in Table 3. 

Table 2 shows the reason why most of the relativistic circles are concentrated at the same region
in spite of the fact that the predicted values of $Q^4_c$ are different from one another. 
The main components of $Q^4_{cn}$ are $Q_{2n}$ and $Q_{4W_n}$. In comparing NL3(5) with NL-SH(6) 
in Table 2, the value of $Q_{2n}$ in the former is larger than that in the latter,
while the value of $Q_{4W_n}$ of NL3 is smaller than that of NL-SH. As a result,
the values for $Q^4_{cn}$ given by their sum are almost the same, but their $Q^4_c$ is
dominated by $Q^4_p$ which has different values in the two models as in Table 2.
The values of $Q^4_p$ depend on $R^2_n$ through Eq.(\ref{qprp}) and(\ref{rprn})
Thus, the spin-orbit density plays an important role in the relativistic models.  
This fact will be seen again in the next subsection.

Similar distributions are obtained between $Q^4_{cn}$ and $Q^4_c$ in $^{48}$Ca and $^{208}$Pb,
as in Figure \ref{QcnQc-Ca40}. Most of the predicted values are concentrated around $Q^4_{cn}=$
12.9 fm$^4$ in $^{48}$Ca and around 33.0 fm$^4$ in $^{208}$Pb. Thus, it is not necessary for the LSA
to provide the linear relationship defined well between physical quantities.
In contrast to the relativistic models, the non-relativistic models predict the values of $Q^4_{cn}$
rather well on the LSL as in Figure \ref{QcnQc-Ca40}, in spite of the fact that the value of $\sigma$
listed in Table 3 is not small. In Appendix, the LSA between $Q^4_{cn}$ and $Q^4_c$ will be discussed
in detail in terms of the correlation coefficients numerically.

\subsection{The rms of the proton
  and neutron densities in $^{48}$Ca}\label{ca48}

\begin{figure}[h]
\begin{minipage}{7.7cm}
\includegraphics[bb=0 0 216 162]{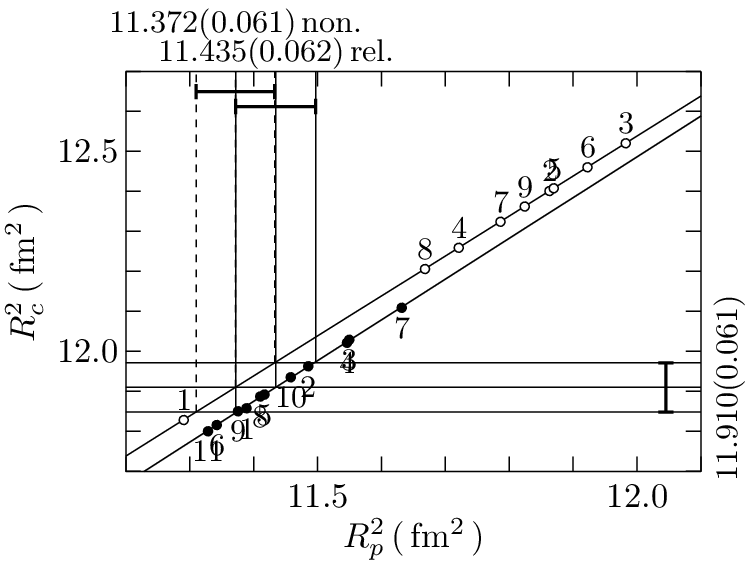}
\caption{The same as Figure \ref{RpRc-Ca40}, but for $R^2_c$ against $R^2_p$
in $^{48}$Ca. 
}
\label{RpRc-Ca48}
\end{minipage}\hspace{0.3cm}%
\begin{minipage}{7.7cm}
\includegraphics[bb=0 0 216 162]{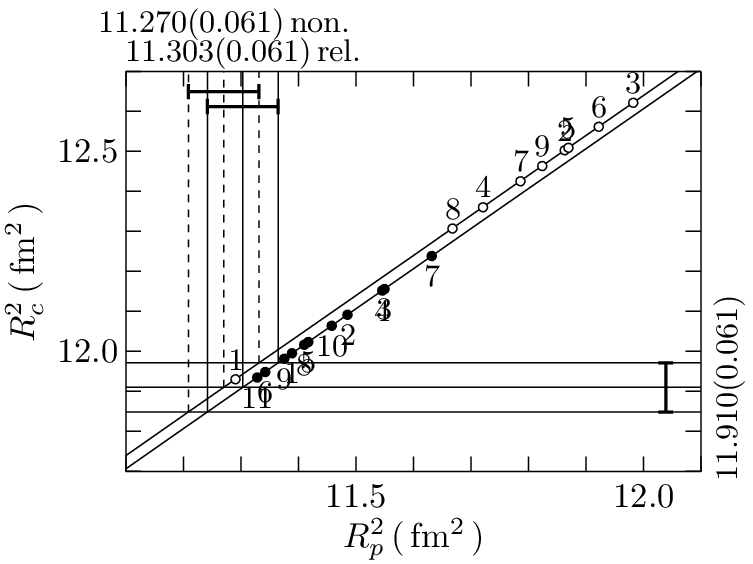}
\caption{The same as Figure \ref{RpRc-Ca40}, but for $R^2_c$
calculated without the spin-orbit density against $R^2_p$
in $^{48}$Ca. For details, see the text.
}
\label{RpRc-Ca48-noLS}
\end{minipage}
\end{figure}

Figure \ref{RpRc-Ca48} shows the LSL's for $R_p^2$ and $R_c^2$
in $^{48}$Ca.
The equations of the lines are listed in Table 4 at the end of this subsection.
The relativistic line provides the accepted region of $R_p^2$ to be 11.435(0.060) fm$^2$,
while the non-relativistic one 
yields that of $R_{p,{\rm nr}}^2$ to be 11.372(0.061) fm$^2$.
Both models have small values of $\sigma$, but all the calculated values
of $R^2_{c, {\rm nr}}$ in the non-relativistic models are outside of the
accepted region, as in the case of $^{40}$Ca in Figure \ref{RpRc-Ca40}.

The difference between the two lines is partially
due to the spin-orbit density which contributes to $R_p^2$ of $^{48}$Ca
in both relativistic and non-relativistic models, but in a different way. 
Figure \ref{RpRc-Ca48-noLS} shows how the contributions are.  
If the spin-orbit density is neglected, the slopes of the both lines are
given by 1.0000, and the difference between their intercepts is almost
equal to the value of $3/4M^2=0.0331$ fm$^2$, as seen in Table 4.

\begin{figure}[h]
\begin{minipage}{7.7cm}
\includegraphics[bb=0 0 211 162]{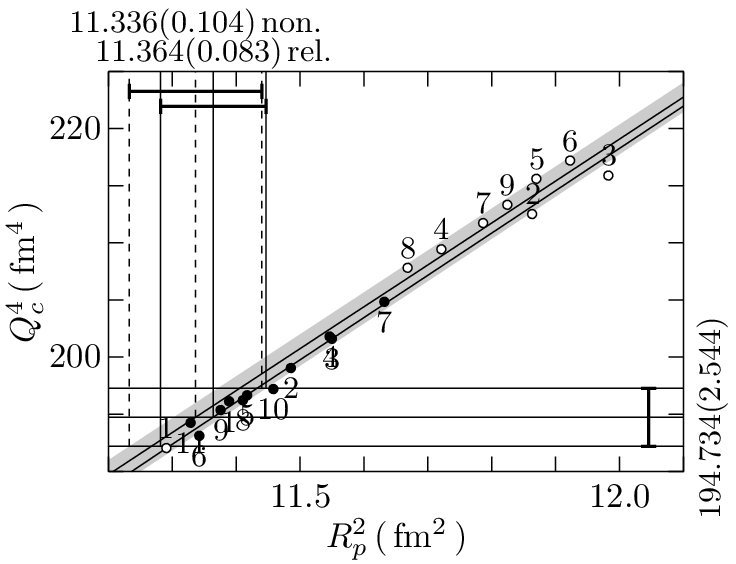}
\caption{The same as Figure \ref{RpRc-Ca40}, but for $Q^4_c$ against $R^2_p$
in $^{48}$Ca.The gray area denotes the standard deviation
of the calculated values from the least square lines.
}
\label{RpQc-Ca48}
\end{minipage}\hspace{0.3cm}
\begin{minipage}{7.7cm}
\includegraphics[bb=0 0 211 162]{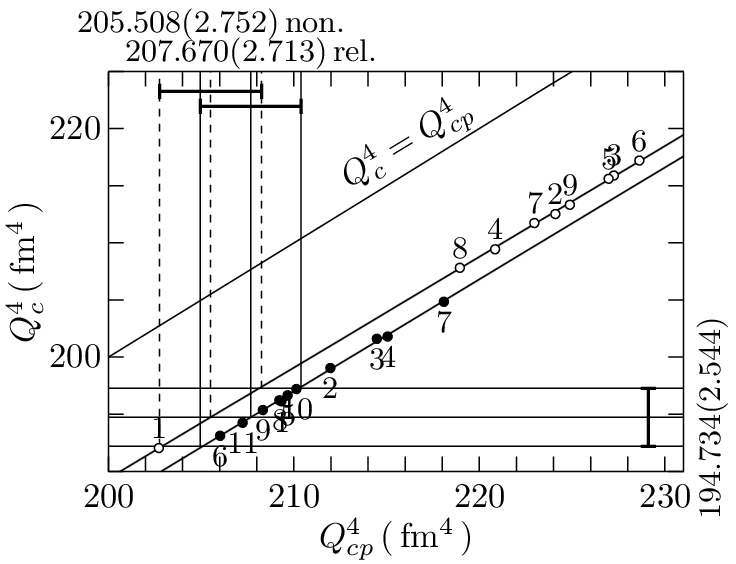}
\caption{The same as Figure \ref{RpRc-Ca40}, but for $Q^4_c$ against
$Q_{cp}^4$ in $^{48}$Ca. The line indicated by $Q^4_c=Q^4_{cp}$
is obtained neglecting the contribution from $Q^4_{cn}$ to $Q^4_c$.
}
\label{QcpQc-Ca48}
\end{minipage}
\end{figure}

Figure \ref{RpQc-Ca48} shows the LSL's between $R_p^2$ and $Q_c^4$.
The two lines for the relativistic and non-relativistic models cross
the experimental region at 11.364(0.069) and 11.336(0.069) fm$^2$, respectively.

Figure \ref{QcpQc-Ca48} shows the LSL's for $Q^4_{cp}$ and $Q^4_c$.
Both relativistic and non-relativistic models predict their values almost on the 
straight lines.
They provide the value of $Q^4_{cp}$ to be 207.670(2.597) and 205.508(2.624)
fm$^4$ for relativistic and non-relativistic frameworks, respectively.
The difference between the two lines is mainly due to  
$Q^4_p$ and $Q_{4W_p}$, as in Figure  \ref{QcpQc-Ca40} for $^{40}$Ca.

If there were no contribution from the neutrons, $Q^4_{cn}$, to $Q^4_c$,
one would have the line which is indicated by $Q^4_c=Q^4_{cp}$ in
Figure \ref{QcpQc-Ca48}. 
The difference between this line and the two LSL's shows the contribution
from $Q^4_{cn}$ to $Q^4_c$. Although the ratio of $Q^4_{cn}$ to $Q^4_c$ is about 5\%,
it is seen that the $Q^4_c=Q^4_{cp}$ line definitely requires the negative contribution
from the neutrons. It should be also noticed that
the line of $Q^4_c=Q^4_{cn}$ is almost parallel to the two LSL's.
This fact implies that the values of $Q^4_{cn}$ are almost
independent of those of $Q^4_c$, and as a result, the LSA
between $Q^4_{cn}$ and $Q^4_c$ is not useful for estimation of  the values
of $Q^4_{cn}$ in $^{48}$Ca, as in the case of $^{40}$Ca.

\begin{figure}[ht]
\begin{minipage}{7.8cm}
\includegraphics[bb=0 0 219 161]{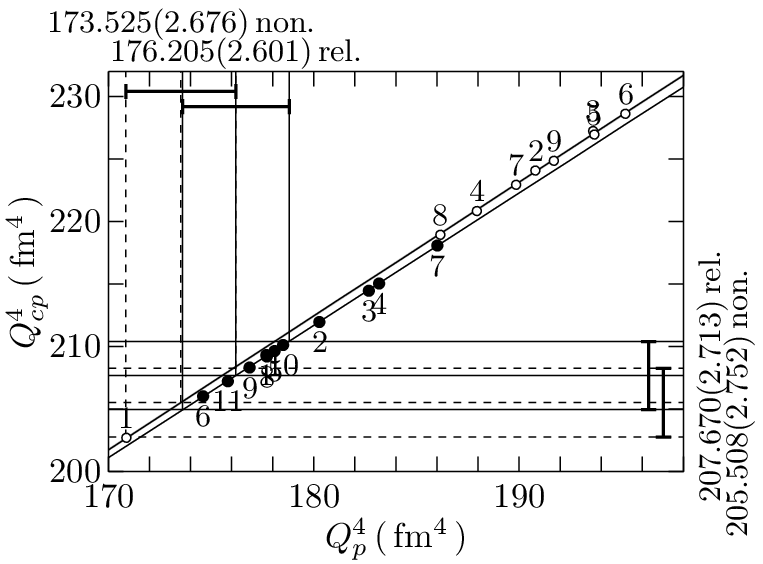}
\caption{The same as Figure \ref{RpRc-Ca40}, but for $Q^4_{cp}$ against $Q^4_p$
in $^{48}$Ca.
}
\label{QpQcp-Ca48}
\end{minipage}\hspace{0.3cm}
\begin{minipage}{7.8cm}
\includegraphics[bb=0 0 219 162]{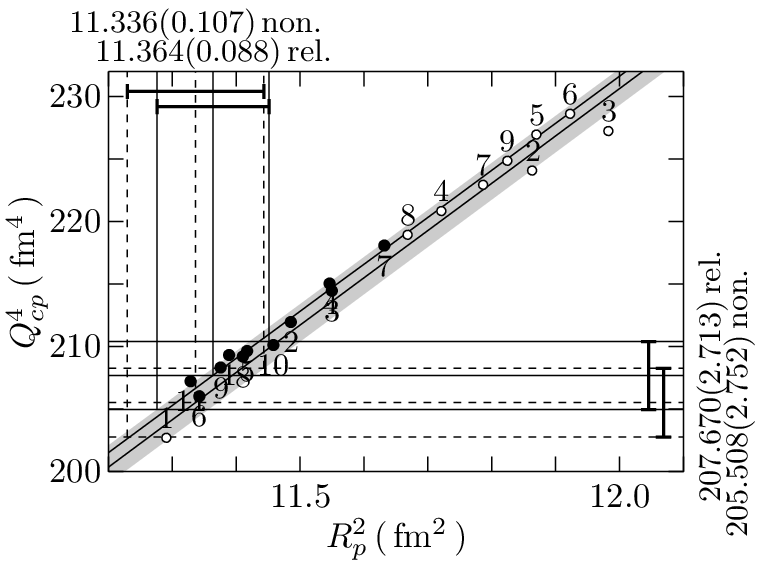}
\caption{The same as Figure \ref{RpRc-Ca40}, but for $Q^4_{cp}$ against $R^2_p$
in $^{48}$Ca. The gray area denotes the standard deviation
of the calculated values from the least square lines.
}
\label{RpQcp-Ca48}
\end{minipage}
\end{figure}

The analysis between $Q^4_p$ and
$Q^4_{cp}$ in Figure \ref{QpQcp-Ca48} shows that the accepted value of $Q^4_p$ is
176.205(2.453) in the relativistic framework, while 173.525(2.452) fm$^4$ in the
non-relativistic one. Most of the relativistic models predict the values in the accepted
region of $Q^4_{cp}$, but all the values in the non-relativistic models  
are outside of the accepted region as in Figure \ref{QpQcp-Ca40} for $^{40}$Ca,
and, in particular, the one of $Q^4_p$ by SLy4(6) is the largest among them.

Using the above accepted values of $Q^4_{cp}$ in $^{48}$Ca,
the LSL's are obtained for $R^2_p$  as in Figure \ref{RpQcp-Ca48}.
The accepted values of $R^2_p$ are the same as those
in Figure  \ref{RpQc-Ca48} for $R_p^2$ and $Q_c^4$.
The reason why they are the same will be seen in Appendix.

Finally from Figures \ref{RpRc-Ca48},  \ref{RpQc-Ca48} and \ref{RpQcp-Ca48}, 
the common accepted region of $R^2_p$ in the relativistic framework is decided to be
$\mathcal{R}_p=11.375 \sim 11.433$ fm$^2$, which corresponds to $R_p= 3.373 \sim 3.381$ fm.
The lower bound is obtained from Figure \ref{RpRc-Ca48}
and the upper bound from Figure \ref{RpQc-Ca48} and \ref{RpQcp-Ca48}.
For the non-relativistic models, they are obtained to be 
$\mathcal{R}_{p, {\rm nr}}=11.311 \sim 11.405$ fm$^2$, yielding $R_{p,{\rm nr}} =3.363 \sim 3.377$ fm.
The lower and the upper bound are from Figure \ref{RpRc-Ca48}, and
\ref{RpQc-Ca48} and \ref{RpQcp-Ca48}, respectively.

\begin{figure}[ht]
\centering{\includegraphics[bb=0 0 216 152]{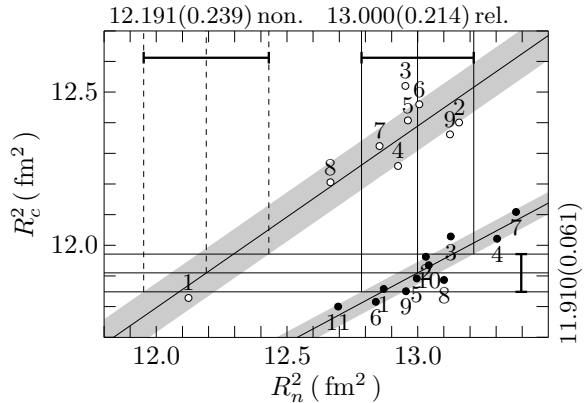}}
\caption{The same as Figure \ref{RpRc-Ca40}, but for $R^2_c$ against $R^2_n$
in $^{48}$Ca. The gray area denotes the standard deviation
of the calculated values from the least square lines.
}
\label{RnRc-Ca48}
\end{figure}

A similar analyses to the one for $R_n^2$ of $^{40}$Ca  are performed for $^{48}$Ca.
Figure \ref{RnRc-Ca48} shows the LSL's for $R_n^2$ and $R_c^2$.
The line for relativistic modes is given by 
$R_c^2=0.4562R_n^2+5.9795$,
and for non-relativistic ones  $R_c^2=0.5922R_n^2+4.6912$.
Unlike the case of $^{40}$Ca, the coefficients of $R^2_n$ are
smaller than 1, owing to the excess neutrons in $^{48}$Ca.
The two lines are much more separated than in Figure \ref{RnRc-Ca40}.
The calculated values expressed by the open and closed circles are
distributed over a similar region of $R^2_n$ around 13 fm$^2$, but, except for SKI(1),
the non-relativistic models overestimate the experimental value of $R^2_c$.
Hence, the LSL of the non-relativistic models yields the smaller value
of $R^2_{n,{\rm nr}}$ to be 12.191(0.104) fm$^2$ than 13.000(0.135) fm$^2$
for the relativistic models.
It should be noticed that if the average value of $R^2_{n,{\rm nr}}$
calculated in the non-relativistic models were compared with that in the relativistic
models, there would be almost no difference between them, 
in contrast to the result of the LSA, as seen in Figure \ref{RnRc-Ca48}.

\begin{figure}[ht]
\centering{\includegraphics[bb=0 0 211 152]{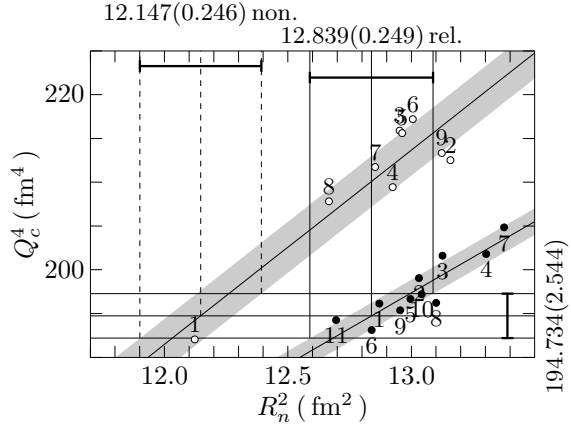}}
\caption{The same as Figure \ref{RpRc-Ca40}, but for $Q^4_c$ against $R^2_n$
in $^{48}$Ca. The gray area denotes the standard deviation
of the calculated values from the least square lines.
}
\label{RnQc-Ca48}
\end{figure}

Figure \ref{RnQc-Ca48} shows the relationship between $R_n^2$ and $Q_c^4$,
which is also expressed by the two lines for the relativistic
and non-relativistic frameworks, respectively.
The accepted region of $R_n^2$ is obtained to be
12.839(0.156) fm$^2$ for the relativistic models, and 12.147(0.115) fm$^2$
for non-relativistic models.
In both Figure \ref{RnRc-Ca48} and \ref{RnQc-Ca48}, the accepted
region of $R_n^2$ from the experimental values are broader than of $R_p^2$
in Figure \ref{RpRc-Ca48} and \ref{RpQc-Ca48},
owing to the difference between the gradients of the LSL's.
As seen in Table 4, the values of the slopes in Figure \ref{RnRc-Ca48}
and \ref{RnQc-Ca48} are smaller than those in Figure \ref{RpRc-Ca48} and
\ref{RpQc-Ca48}, respectively. This is because of $R^2_n > R^2_p$.
For example, the equations of the LSL's in Figure \ref{RpRc-Ca48}
and \ref{RnRc-Ca48} provide
\[
 R^2_p=0.4475R^2_n+5.6171,
\]
in the relativistic models.

\begin{figure}[ht]
\begin{minipage}{7.7cm}
\includegraphics[bb=0 0 206 161]{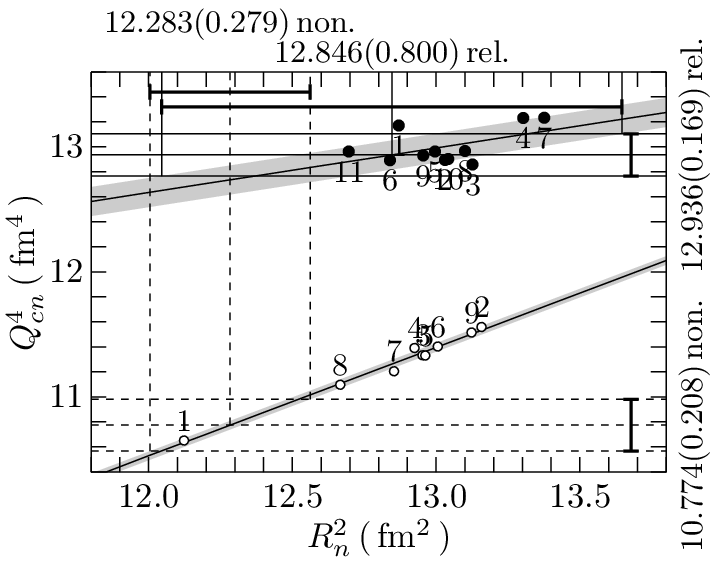}
\caption{The same as Figure \ref{RpRc-Ca40}, but for $Q^4_{cn}$ against $R^2_n$
in $^{48}$Ca. The gray area denotes the standard deviation
of the calculated values from the least square lines.
}
\label{RnQcn-Ca48}
\end{minipage}\hspace{0.3cm}
\begin{minipage}{7.7cm}
\includegraphics[bb=0 0 206 152]{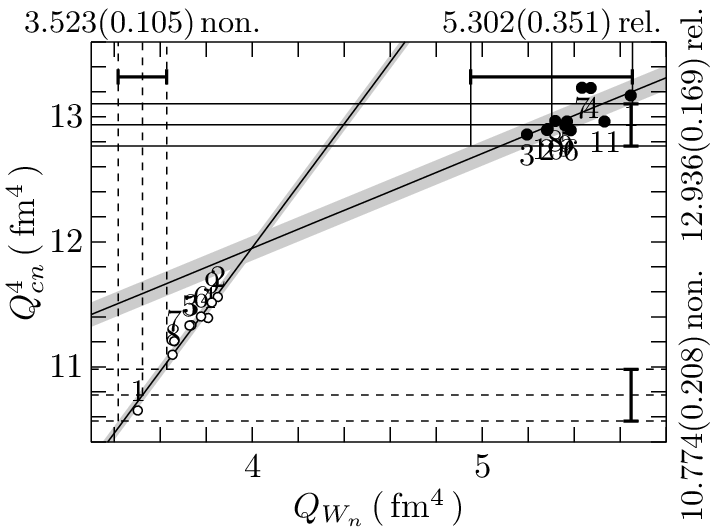}
\caption{The same as Figure \ref{RpRc-Ca40}, but for $Q^4_{cn}$ against
$Q_{Wn}(Q_{2W_n}+Q_{4W_n})$ in $^{48}$Ca. For details, see the text.
}
\label{QWnQcn-Ca48}
\end{minipage}
\end{figure}

Figure \ref{QcpQc-Ca48} provides the accepted values of $Q^4_{cn}$
to be 12.936(0.053) and 10.774(0.080) fm$^4$ for the relativistic
and non-relativistic frameworks, respectively, neglecting $\sigma$.
The reason of the difference between these values is the same as in $^{40}$Ca.
Using these values for $Q^4_{cn}$, Figure \ref{RnQcn-Ca48} determines
the accepted regions of $R_n^2$ by the relationship between $R^2_n$
and  $Q^4_{cn}$. The relativistic line is given by $Q_{cn}^4=0.3560R_n^2+8.3633$,
and the non-relativistic one by $Q_{cn}^4=0.8678R_n^2+0.1149$.
The calculated values of $Q^4_{cn,\,{\rm nr}}$ in the non-relativistic models
are well on the line,
while the relativistic ones are distributed around the line,
although most of them are within the experimental error.
The reason of this fact is understood in a similar way as for Figure \ref{RnQcn-Ca40}
of $^{40}$Ca, but will be discussed in more detail as follows.

As seen in Eq.(\ref{4thm}),  $Q^4_{cn}$ is composed of the four terms.
Among them, $Q_{2n}$ and $Q_{4W_n}$ are responsible for the change of
$Q^4_{cn}$ with $R^2_n$.
The former is proportional to $R^2_n$. In the non-relativistic models,
the latter is given by the radius of the neutrons
in the $f_{7/2}$ shell, according to Eq.(\ref{wso}).
It is expected that its radius also increases with increasing $R^2_{n,{\rm nr}}$.
This fact is seen in Figure \ref{QWnQcn-Ca48} which shows the LSL
for $Q_{W_n}$ and $Q^4_{cn}$, where $Q_{W_n}=Q_{2W_n}+Q_{4W_n}$,
but the values of $Q_{2Wn}$ is small, compared with that of $Q_{4Wn}$.
In Figure \ref{QWnQcn-Ca48} and Figure \ref{RnQcn-Ca48}, it is seen that 
the number indicating each model is in the same order on the LSL's
of the non-relativistic models.
In the relativistic case, there is no such a correlation
between the numbers in Figure \ref{QWnQcn-Ca48} and Figure \ref{RnQcn-Ca48},
since Eq.(\ref{wso}) does not hold.
This fact is also the reason in Figure \ref{RnQcn-Ca48}
why most of the relativistic models predict the values within the band of
$Q^4_{cn}$, 12.936(0.169) fm$^2$, in spite of the fact that their values
of $R_n^2$ are different from one another. Among the relativistic models,
the one which predicts a smaller value of $R_n^2$ yields a larger value
of $Q_{4W_n}$, and vice versa. For example,
in Figure \ref{RnQcn-Ca48}, FSU(11) predicts the smallest
$R_n^2$, while on the contrary, in Fig \ref{QWnQcn-Ca48}
it provides the largest value of the spin-orbit contribution
within the band of the experimental value.

Figure \ref{RnQcn-Ca48} shows that the accepted region of $R_n^2$ from the
lines between $R_n^2$ and $Q^4_{cn}$ are given as
12.846(0.150) and 12.283(0.093) for relativistic and non-relativistic frameworks,
respectively.

The common accepted region of $R^2_n$ is obtained from Figure 
\ref{RnRc-Ca48}, \ref{RnQc-Ca48}
and \ref{RnQcn-Ca48}, neglecting $\sigma$, as follows.
In relativistic models, $\mathcal{R}_n=12.865\sim 12.995$ fm$^2$,
corresponding to $R_n=3.587\sim 3.605$ fm.
The lower bound is given by Figure \ref{RnRc-Ca48},
and the upper bound by Figure \ref{RnQc-Ca48}.
These values are not affected by Figure \ref{RnQcn-Ca48},
since its accepted region contains the above one from 12.865 to 12.995 fm$^2$.
In non-relativistic models, Figure \ref{RnQcn-Ca48} provides the lower bound,
and Figure \ref{RnQc-Ca48} the upper bound as $\mathcal{R}_{n, {\rm nr}}=12.190\sim 12.262$ 
fm$^2$, yielding $R_{n,{\rm rn}} =3.491\sim 3.502$ fm.
These region are contained in the accepted region by Figure \ref{RnRc-Ca48}.
In $^{48}$Ca, the lower bound is determined by the relationship
between $R^2_{n,{\rm nr}}$ and $Q^4_{cn,{\rm nr}}$.

From the above results for $R_n$ and $R_p$, the skin thickness defined by
$\delta R=R_n-R_p$ is given to be $0.206\sim 0.232$ fm
in the relativistic framework, and $0.114\sim 0.139$ fm in the
non-relativistic framework. 
The skin thickness of $^{48}$Ca in the relativistic models is larger
by 0.067$\sim$0.118 fm than in the non-relativistic models.
The difference between $\delta R$ mainly stems from the rms
of neutron distributions in the two models.

\begin{table}
\renewcommand{\arraystretch}{1.3}
\begin{tabular}{|c|c|c||r|r|r||r|r|r|} \hline
\multicolumn{3}{|c||}{$^{48}$Ca}  & \multicolumn{3}{c||}{Rel.} &
\multicolumn{3}{c|}{Non.} \\ \hline
Fig.  & $y$     & $x$     & \multicolumn{1}{c|}{$a$} & \multicolumn{1}{c|}{$b$} &
\multicolumn{1}{c||}{$\sigma$} & \multicolumn{1}{c|}{$a$} & \multicolumn{1}{c|}{$b$} &
\multicolumn{1}{c|}{$\sigma$} \\ \hline
\ref{RpRc-Ca48} &  $R_c^2$ & $R_p^2$ &
$ 1.0195$ & $    0.2529$ & $ 0.0023$ & $ 1.0000$ & $    0.5385$ & $ 0.0000$ \\ \hline
\ref{RpRc-Ca48-noLS} &  $R_c^2$  & $R_p^2$  &
$ 1.0000$ & $    0.6067$ & $ 0.0000$ & $ 1.0000$ & $    0.6399$ & $ 0.0000$ \\ \hline
\ref{RpQc-Ca48} & $Q_c^4$ & $R_p^2$  & 
$37.0126$ & $ -225.8796$ & $ 0.5114$ & $36.7018$ & $ -221.3314$ & $ 1.2683$ \\ \hline
\ref{QcpQc-Ca48}& $Q_c^4$ &  $Q_{cp}^4$  &
$ 0.9794$ & $   -8.6631$ & $ 0.1131$ & $ 0.9693$ & $   -4.4715$ & $ 0.1236$ \\ \hline
\ref{QpQcp-Ca48}& $Q_{cp}^4$ &  $Q_{p}^4$  &
 $ 1.0587$ & $   21.1178$ & $ 0.0412$ & $ 1.0701$ & $   19.8141$ & $ 0.1122$ \\ \hline
\ref{RpQcp-Ca48}&  $Q_{cp}^4$  &  $R_{p}^2$  &
$37.6573$ & $ -220.2584$ & $ 0.5857$ & $37.8665$ & $ -223.7615$ & $ 1.2985$ \\ \hline
\ref{RnRc-Ca48}&  $R_{c}^2$  &  $R_{n}^2$  &
$ 0.4562$ & $    5.9795$ & $ 0.0364$ & $ 0.5922$ & $    4.6912$ & $ 0.0802$ \\ \hline
\ref{RnQc-Ca48}&  $Q_{c}^4$  &  $R_{n}^2$  &
$16.2922$ & $  -14.4378$ & $ 1.5207$ & $22.1955$ & $  -74.8670$ & $ 2.9163$ \\ \hline
\ref{RnQcn-Ca48}&  $Q_{cn}^4$  &  $R_{n}^2$  &
$ 0.3560$ & $    8.3633$ & $ 0.1160$ & $ 0.8678$ & $    0.1149$ & $ 0.0339$ \\ \hline
\ref{QWnQcn-Ca48} &  $Q_{cn}^4$  & $Q_{W_n}$  &
$ 0.7581$ & $    8.9166$ & $ 0.0975$ & $ 2.4751$ & $    2.0539$ & $ 0.0531$ \\ \hline
\end{tabular}
\caption{The least square line $y(x)=ax+b$ and the standard deviation $\sigma$ depicted
 in Figure \ref{RpRc-Ca48} to \ref{QWnQcn-Ca48} for the relativistic(Rel.)
 and the non-relativistic(Non.) models.
 }
\end{table}

Finally one comment is added in this subsection.
On the one hand, in $^{40}$Ca, the difference between $R^2_c$ and $R^2_p$ obtained
in the LSA, $R^2_c-R^2_p=11.903-11.252(11.216)=0.651(0.687)$ fm$^2$
in the relativistic(non-relativistic) models,
neglecting the error, is almost equal to
the contributions to the msr from the nucleon form factors,
$r^2_p+(r^2_+-r^2_-)=0.653$ fm$^2$.
On the other hand, in $^{48}$Ca, 
the difference between $R^2_c$ and $R^2_p$ is given by
$R^2_c-R^2_p=11.910-11.435(11.372)=0.475(0.538)$ fm$^2$
in the relativistic(non-relativistic) models which is smaller
by 0.178(0.115) fm$^2$ than that from the contribution of the
nucleon form factor. This negative contribution in $^{48}$Ca stems from the
spin-orbit density of the excess neutrons.
The importance of the neutron charge density
is also seen in comparing the experimental value of
$R_c$ in $^{40}$Ca with that in $^{48}$Ca. They are almost the same,
as 3.450(0.010) and 3.451(0.009) fm.
In contrast to this fact, the estimated value of $R_p$ in $^{48}$Ca
is 3.377(0.004) fm in the relativistic models and 3.370(0.007) fm
in the non-relativistic ones, while that in $^{40}$Ca is 3.346(0.002) fm
and 3.346(0.007) fm, respectively. 
The reduction of $^{48}$Ca values is owing to the negative contribution from
the neutron charge density in addition to that from the spin-orbit density.
Moreover, it is understood as the contribution of the neutrons
why the experimental value of $Q_c$ of $^{48}$Ca is smaller than
that of $^{40}$Ca.
These neutron effects should not be disregarded in the
detailed discussions such as on the isotope shift, as in Refs.\cite{nor,ruiz}.

\subsection{The rms of the proton and neutron densities in $^{208}$Pb}\label{pb}

Among stable neutron-rich nuclei, $^{208}$Pb is also appropriate
for investigating the fourth-order moment with the mean field models.
In this subsection, $R_p$ and $R_n$ of $^{208}$Pb will be estimated
in the same way as those for $^{40}$Ca and $^{48}$Ca.
Since the method of the present analysis has been explained in detail
in the previous subsections,
the present subsection will focus mainly on the results of $^{208}$Pb.

\begin{figure}[ht]
\begin{minipage}{7.7cm}
\includegraphics[bb=0 0 216 162]{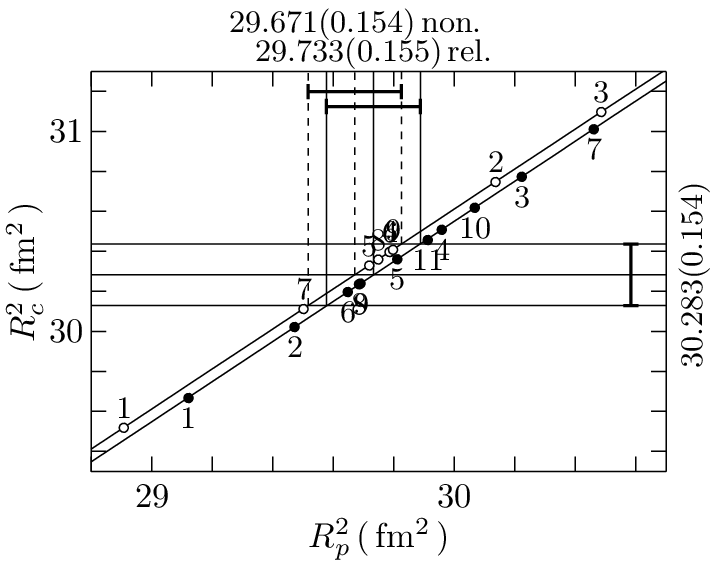}
\caption{The same as Figure \ref{RpRc-Ca40}, but for $R^2_c$ against $R^2_p$
in $^{208}$Pb. 
}
\label{RpRc-Pb208}
\end{minipage}\hspace{0.3cm}
\begin{minipage}{7.7cm}
\includegraphics[bb=0 0 216 162]{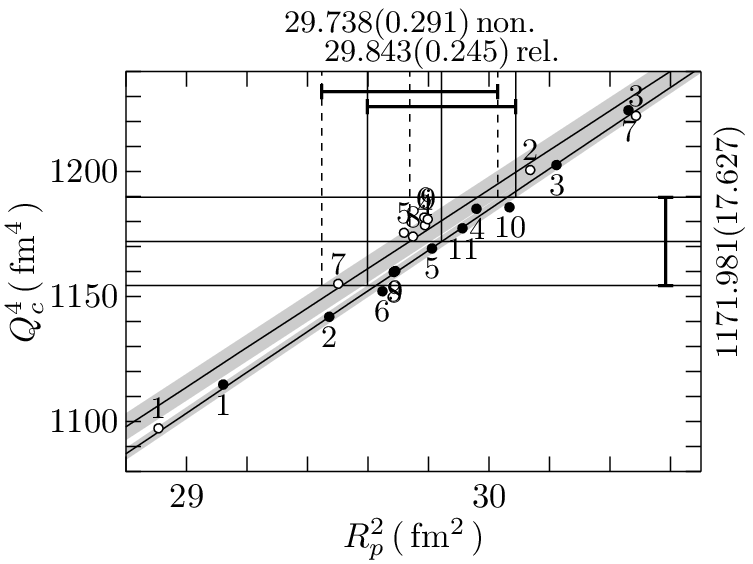}
\caption{The same as Figure \ref{RpRc-Ca40}, but for $Q^4_c$ against $R^2_p$
in $^{208}$Pb. The gray area denotes the standard deviation
of the calculated values from the least square lines.
}
\label{RpQc-Pb208}
\end{minipage}
\end{figure}

Figure \ref{RpRc-Pb208} shows the LSL's for $R_p^2$ and $R^2_c$,
whose equations are tabulated in Table 5 at the end of this subsection.
The relativistic line crosses the band of experimental value at 30.283(0.154) fm$^2$,
yielding the accepted value of $R^2_p$ to be 29.733(0.154) fm$^2$,
while the non-relativistic one yielding 29.671(154) fm$^2$.
It is seen that in the case of $^{208}$Pb, some of the evaluated values
in the non-relativistic models also are within the band
of the experimental value, unlike in the case of Ca isotopes. 

The LSL's for $R^2_p$ and $Q^4_c$ in Figure \ref{RpQc-Pb208} provide the accepted region
of $R^2_p$ to be 29.843(0.216) fm$^2$ and 29.738(0.223) fm$^2$ for relativistic
and non-relativistic frameworks, respectively.

\begin{figure}[ht]
\begin{minipage}{7.7cm}
\includegraphics[bb=0 0 216 162]{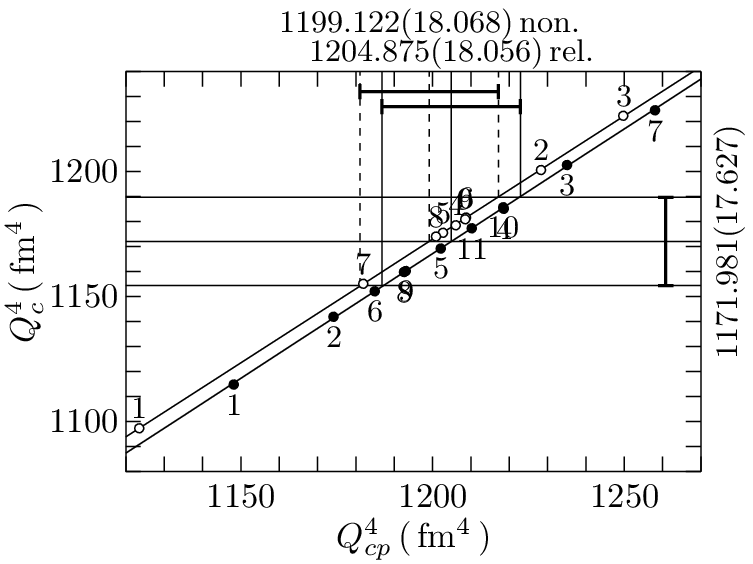}
\caption{The same as Figure \ref{RpRc-Ca40}, but for $Q^4_c$ against $Q^4_{cp}$
in $^{208}$Pb. 
}
\label{QcpQc-Pb208}
\end{minipage}\hspace{0.3cm}
\begin{minipage}{7.9cm}
\includegraphics[bb=0 0 224 162]{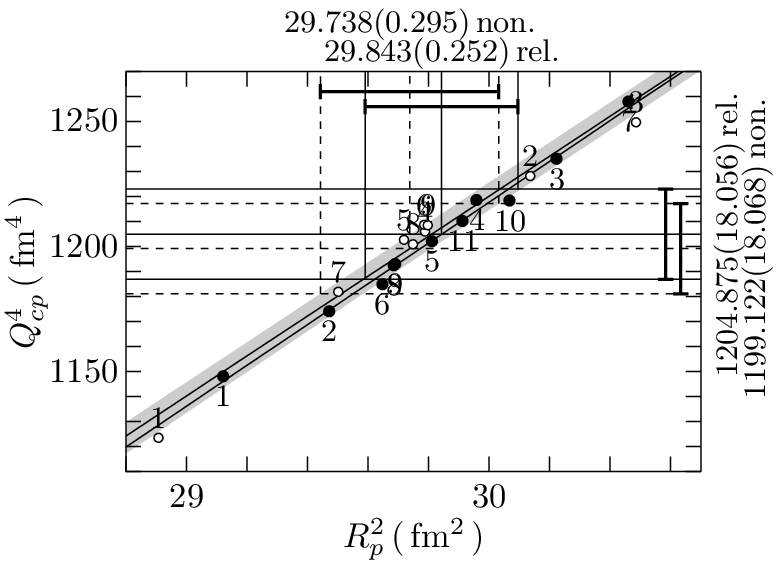}
\caption{The same as Figure \ref{RpRc-Ca40}, but for $Q^4_{cp}$ against $R^2_p$
in $^{208}$Pb. The gray area denotes the standard deviation
of the calculated values from the least square lines.
}
\label{RpQcp-Pb208}
\end{minipage}
\end{figure}

The accepted values of $Q^4_{cp}$ is obtained from the analysis
between $Q^4_{cp}$ and $Q^4_c$ in Figure \ref{QcpQc-Pb208}.
They are 1204.875(17.676) fm$^4$ and 1199.122(17.845) fm$^4$.
Using these values, Figure \ref{RpQcp-Pb208} shows the accepted values of
$R^2_p$, which are almost the same as those in Figure \ref{RpQc-Pb208}.
In neglecting $\sigma$, the corresponding two figures yield 
the same accepted region. The relationship between these two figures
will be discussed in Appendix.

The common accepted regions of $R^2_p$ in Figure 
\ref{RpRc-Pb208}, \ref{RpQc-Pb208} and \ref{RpQcp-Pb208} are given by
$\mathcal{R}_p=29.627\sim 29.887$ fm$^2$ in the relativistic models.
They correspond to $R_p$ to be $5.443 \sim 5.467$ fm.
The lower bound is from Figure \ref{RpQc-Pb208} or \ref{RpQcp-Pb208},
while the upper bound from Figure \ref{RpRc-Pb208}.
For non-relativistic models, both the lower and the upper bound are given by
Figure \ref{RpRc-Pb208} as $\mathcal{R}_{p,{\rm nr}}=29.517 \sim 29.825$ fm$^2$,
giving $R_{p,{\rm nr}}=5.433 \sim 5.461$ fm.

\begin{figure}[ht]
\centering{\includegraphics[bb=0 0 224 161]{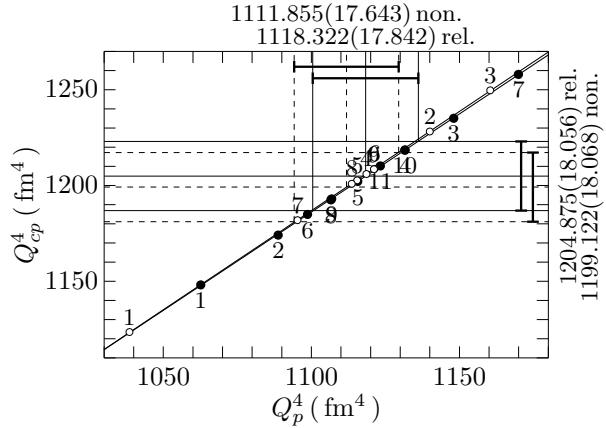}}
\caption{The same as Figure \ref{RpRc-Ca40}, but for $Q^4_{cp}$ against $Q^4_p$
in $^{208}$Pb. 
}
\label{QpQcp-Pb208}
\end{figure}

As a reference, Figure \ref{QpQcp-Pb208} shows the relationship between
$Q^4_p$ and $Q^4_{cp}$ in $^{208}$Pb,
which well determine the accepted values of $Q^4_p$.  
The values of $Q^4_{cp}$ on the right-hand side are taken from Figure \ref{QcpQc-Pb208},
taking account of $\sigma$, but, without $\sigma$, they are 1204.875(17.676) fm$^4$
and 1199.122(17.845) fm$^4$ for the relativistic and non-relativistic models, respectively. 

\begin{figure}[ht]
\begin{minipage}{7.7cm}
\includegraphics[bb=0 0 216 161]{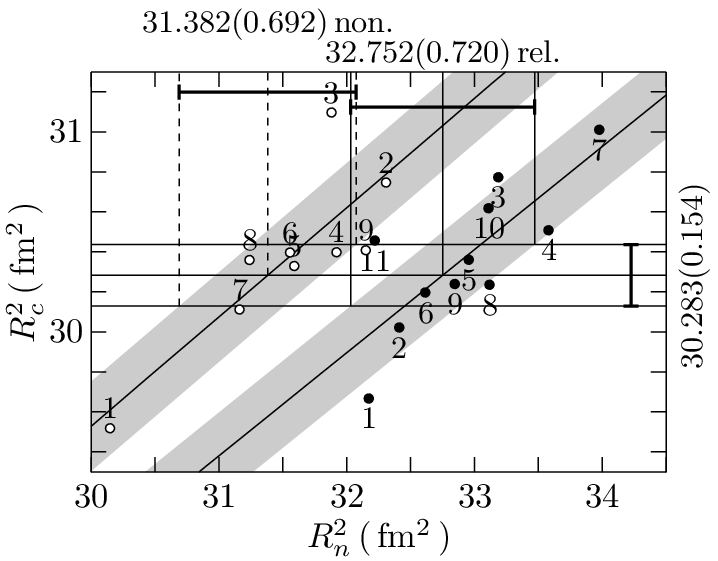}
\caption{The same as Figure \ref{RpRc-Ca40}, but for $R^2_c$ against $R^2_n$
in $^{208}$Pb. The gray area denotes the standard deviation
of the calculated values from the least square lines.
}
\label{RnRc-Pb208}
\end{minipage}\hspace{0.3cm}
\begin{minipage}{7.7cm}
\includegraphics[bb=0 0 216 161]{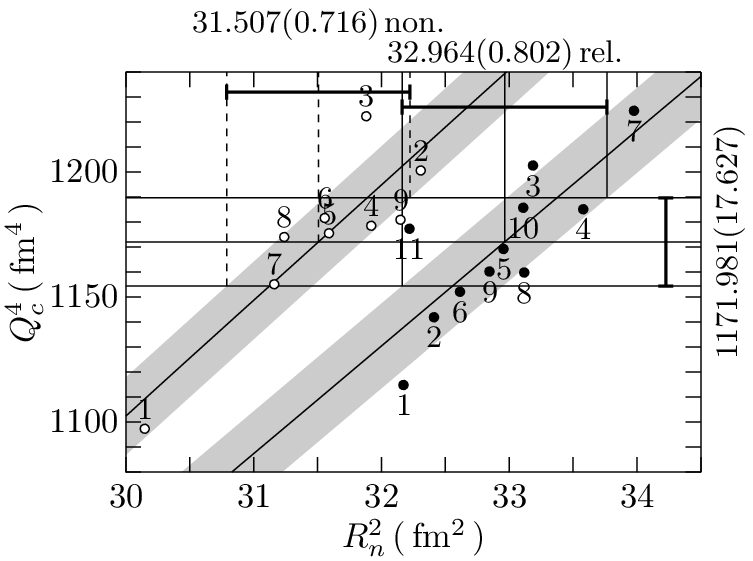}
\caption{The same as Figure \ref{RpRc-Ca40}, but for $Q^4_c$ against $R^2_n$
in $^{208}$Pb. The gray area denotes the standard deviation
of the calculated values from the least square lines.
}
\label{RnQc-Pb208}
\end{minipage}
\end{figure}

\begin{figure}[ht]
\centering{\includegraphics[bb=0 0 206 161]{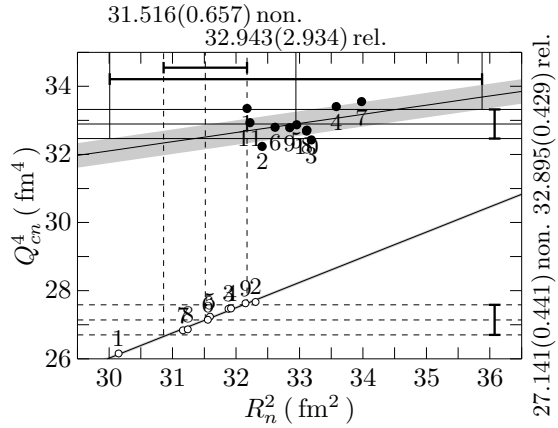}}
\caption{The same as Figure \ref{RpRc-Ca40}, but for $Q^4_{cn}$ against $R^2_n$
in $^{208}$Pb. The gray area denotes the standard deviation
of the calculated values from the least square lines.
}
\label{RnQcn-Pb208}
\end{figure}

The estimation of $R^2_n$ is performed according to Figure 
\ref{RnRc-Pb208}, \ref{RnQc-Pb208} and \ref{RnQcn-Pb208},
which show the LSL's for $R^2_n$ and $R^2_c$, $R^2_n$ and $Q^4_c$,
and $R^2_n$ and $Q^4_{cn}$, respectively.
The accepted values of $Q^4_{cn}$ in the last analysis is obtained
using Eq.(\ref{4thm}) together with the figure without $\sigma$
corresponding to Figure \ref{QcpQc-Pb208}.
They are 32.895(0.049) fm$^4$ and 27.141(0.218) fm$^4$
for relativistic and non-relativistic models, respectively.
Unlike in the cases of $^{40}$Ca and $^{48}$Ca, some of the non-relativistic
models predict the values of $R^2_{n, {\rm nr}}$ within the accepted regions
determined from $R^2_c$, $Q^4_c$ and $Q^4_{cn}$, in a similar way to those
of the relativistic models.
This fact reflects the distribution of the values of $R^2_{p, {\rm nr}}$
in the non-relativistic models in Figure \ref{RpRc-Pb208}, \ref{RpQc-Pb208}
and \ref{RpQcp-Pb208}.

Finally we obtain the common accepted region of $R^2_n$ to be
$\mathcal{R}_n=32.761\sim 33.051$ fm$^2$ in the relativistic framework,
where the lower bound is given by Figure  \ref{RnQcn-Pb208}, and the upper bound
from Figure \ref{RnRc-Pb208}. The corresponding $R_n$ are $5.724\sim 5.749$ fm.
In the non-relativistic framework, we have $\mathcal{R}_{n,{\rm nr}}= 31.221\sim 31.664$ fm$^2$,
which provides $R_{n,{\rm nr}}=5.588\sim 5.627$ fm. The lower bound is provided by
Figure \ref{RnQcn-Pb208}, and the upper bound by Figure \ref{RnRc-Pb208}.

\begin{table}
\renewcommand{\arraystretch}{1.3}
\begin{tabular}{|c|c|c||r|r|r||r|r|r|} \hline
\multicolumn{3}{|c||}{$^{208}$Pb}  & \multicolumn{3}{c||}{Rel.} &
\multicolumn{3}{c|}{Non.} \\ \hline
Fig.  & $y$     & $x$     & \multicolumn{1}{c|}{$a$} & \multicolumn{1}{c|}{$b$} &
\multicolumn{1}{c||}{$\sigma$} & \multicolumn{1}{c|}{$a$} & \multicolumn{1}{c|}{$b$} &
\multicolumn{1}{c|}{$\sigma$} \\ \hline
\ref{RpRc-Pb208} &  $R_c^2$ & $R_p^2$ &
$ 1.0024$ & $    0.4791$ & $ 0.0015$ & $ 1.0000$ & $    0.6116$ & $ 0.0000$ \\ \hline
\ref{RpQc-Pb208} & $Q_c^4$ & $R_p^2$ & 
$81.4556$ & $-1258.9107$ & $ 2.3239$ & $79.0078$ & $-1177.5469$ & $ 5.3272$ \\ \hline
\ref{QcpQc-Pb208}& $Q_c^4$ &  $Q_{cp}^4$ &
$ 0.9972$ & $  -29.5610$ & $ 0.3790$ & $ 0.9878$ & $  -12.4784$ & $ 0.2200$ \\ \hline
\ref{RpQcp-Pb208} & $Q_{cp}^4$ & $R_p^2$ & 
$81.6203$ & $-1230.9324$ & $ 2.5499$ & $79.9479$ & $-1178.3627$ & $ 5.4793$ \\ \hline
\ref{QpQcp-Pb208}& $Q_{cp}^4$ &  $Q_{p}^4$ &
$ 1.0258$ & $   57.6888$ & $ 0.2469$ & $ 1.0352$ & $   48.1484$ & $ 0.1958$ \\ \hline
\ref{RnRc-Pb208}&  $R_{c}^2$ &  $R_{n}^2$ &
$ 0.5154$ & $   13.4036$ & $ 0.2170$ & $ 0.5463$ & $   13.1395$ & $ 0.2242$ \\ \hline
\ref{RnQc-Pb208}&  $Q_{c}^4$ &  $R_{n}^2$ &
$43.1093$ & $ -249.0948$ & $16.9393$ & $46.2100$ & $ -283.9406$ & $15.4732$ \\ \hline
\ref{RnQcn-Pb208}&  $Q_{cn}^4$ &  $R_{n}^2$ &
$ 0.2689$ & $   24.0364$ & $ 0.3599$ & $ 0.7399$ & $    3.8218$ & $ 0.0450$ \\ \hline
\end{tabular}
\caption{The least square line $y(x)=ax+b$ and the standard deviation $\sigma$ depicted
 in Figure \ref{RpRc-Pb208} to \ref{RnQcn-Pb208} for the relativistic(Rel.)
 and the non-relativistic(Non.) models. }
\end{table}

According to the obtained values of $R_n$ and $R_p$, the skin thickness
of $^{208}$Pb is determined to be 0.257$\sim$0.306 fm in the relativistic
models, while 0.127$\sim$ 0.194 fm in the non-relativistic ones.

It is known, for example, as shown in Ref.\cite{roca},
that the predicted value of $R_n$ in $^{208}$Pb is larger
in relativistic models than in non-relativistic models.
In the present analysis, the values of the relativistic models are larger
by $\sim 0.1$ fm than those of the non-relativistic models in both $^{208}$Pb
and $^{48}$Ca, in spite of the fact that most of the relativistic models reproduce the
experimental values of $R_p$ for both nuclei,
but the non-relativistic models fail to explain them for $^{48}$Ca,
as shown in Figure \ref{RpRc-Ca48}.
This result reflects definitely some difference between the structures
of the two mean field models.
It should be investigated what causes the 0.1 fm difference and
whether or not the difference is avoidable\cite{ksp}, since it is not
a small amount for various problems\cite{thi,roca,roca2}

 It may be useful to compare the present results with those from the
 analyses of the experimental data by hadronic probes summarized
 in Ref.\cite{thi}.

\section{Summary}\label{sm}

According to the least squares analysis (LSA) with respect to the various moments
of the nuclear density in the mean field models\cite{abra,roca},
the mean square radii(msr) of the point proton($R^2_p$) and neutron($R^2_n$)
densities in $^{40}$Ca, $^{48}$Ca and $^{208}$Pb are estimated
with use of the experimental values of the second($R^2_c$)- and
fourth($Q^4_c$)-order moments of the charge densities.
Those experimental values have been determined
through electron scattering\cite{vries,emrich},
where the reaction mechanism and the interaction between the electron
and the nucleus are well known\cite{deforest,bd}.
The structure of the observed electromagnetic moments also is well
understood on the same relativistic basis.
Unlike the conventional analysis for deriving $R_c$ in electron scattering, however,
the LSA is not for determination of the experimental values of $R_p$ and $R_n$
model-independently. It provides the employed model-framework with the values
of $R_p$ and $R_n$  which are consistent with experiment. 
If there is another framework, it may yield a different least square line(LSL),
so that a different value of $R_p$ or $R_n$ would be obtained
for the relevant framework, as in the present paper for the relativistic
and non-relativistic mean field models.

The analyses are performed on the basis of the relationship between various moments
of the proton and the neutron density evaluated by the 11 relativistic
and 9 non-relativistic mean field models.
They are arbitrarily chosen among more than 100 versions
of the parameterizations for their phenomenological
nuclear interactions developed for several decades\cite{sw1,fsu,sw,stone}.
The LSA has been possible only after those 40 years accumulation\cite{roca}.

The msr of the charge density($R^2_c$) is dominated by  
$R^2_p$, while $Q^4_c$ depends on $R^2_n$ also.
Moreover,  $R^2_n$ is implicitly not independent of $R^2_c$ in the nuclear models,
since they are strongly correlated with each other through nuclear interactions. 
Employing these facts, the LSL's are obtained, and
their intersection points with the lines for the experimental values
of $R^2_c$ and $Q^4_c$ are used to determine the values of $R^2_p$
and $R^2_n$ accepted in the mean field models.

For these purpose, it is necessary to have both the relativistic
and the non-relativistic expressions of $R^2_c$ and $Q^4_c$
as exactly and consistently as possible.
Except for the non-relativistic expression of $Q^4_c$,
those have been given in Ref.\cite{ksmsr}. The non-relativistic
one of $Q^4_c$ is derived in the present paper with the help
of the Foldy-Wouthuysen transformation, following Ref.\cite{ksmsr}.
In the definition of $R^2_c$ and $Q^4_c$, the center of mass corrections are ignored.

All the results of the present paper are summarized in Table 6 and 7 in units of fm.
The results in Table 6 are obtained in the analysis of the previous section,
taking account of the experimental errors, but neglecting the standard deviations $\sigma$
of the LSL's listed in Table 3 to 5.
Table 7 shows the results by the analysis taking into account $\sigma$ also.
The way to take account of $\sigma$ is explained in Appendix.

In these tables, the difference between $R_n$ and $R_p$ is given by $\delta R=R_n-R_p$.
The present analyses yield the values of the mean fourth-order moment
of the point($Q^4_p$)and charge($Q^4_{cp}$) proton densities,
and that of the neutron charge density($Q^4_{cn}$) also, as listed in the same tables.
The values of $Q^4_{cp}$ are determined through $Q^4_c$ from the LSA
in the same way as for $Q^4_p$, and those of $Q^4_{cn}$ are obtained by the definition,
$Q^4_c=Q^4_{cp}-Q^4_{cn}$.
In Table 6, the numbers in the parentheses indicate the errors stemming from the
experiment\cite{vries, emrich}, while in Table 7, those contain the errors
coming from $\sigma$ also.
Those errors are less than $\pm 0.5\%$ in Table 6, and less than $\pm 1.0\%$ in Table 7,
compared with their central values, except for the ones of $\delta R$.

\begin{table}
\begingroup
\renewcommand{\arraystretch}{1.2}
\hspace*{-1cm}%
{\setlength{\tabcolsep}{4pt}
\begin{tabular}{|c|l|l|l|l|l|l|l|} \hline
          &      &
\multicolumn{1}{c|}{$R_p$} &
\multicolumn{1}{c|}{$R_n$} &
\multicolumn{1}{c|}{$\delta R$} &
\multicolumn{1}{c|}{$Q_p$} &
\multicolumn{1}{c|}{$Q_{cp}$} & 
\multicolumn{1}{c|}{$Q_{cn}$} \\ \hline
           & Rel. & $3.346(0.002)$ & $3.296(0.002)$ & $-0.050(0.004)$ & $3.635(0.015)$ & $3.785(0.014)$ & $1.513(0.001)$ \\
$^{40}$Ca  & Non. & $3.346(0.007)$ & $3.302(0.006)$ & $-0.045(0.013)$ & $3.628(0.015)$ & $3.782(0.014)$ & $1.466(0.002)$ \\ \cline{2-8}
           & Exp. & \multicolumn{3}{c|}{$R_c=3.450(0.010)$} & \multicolumn{3}{c|}{$Q_c=3.761(0.014)$} \\ \hline
           & Rel. & $3.377(0.004)$ & $3.596(0.009)$  & $0.219(0.013)$ & $3.643(0.013)$ & $3.796(0.012)$ & $1.897(0.002)$ \\
$^{48}$Ca  & Non. & $3.370(0.007)$ & $3.497(0.005)$  & $0.127(0.012)$ & $3.629(0.013)$ & $3.786(0.012)$ & $1.812(0.004)$ \\ \cline{2-8}
           & Exp. & \multicolumn{3}{c|}{$R_c=3.451(0.009)$} & \multicolumn{3}{c|}{$Q_c=3.736(0.012)$} \\ \hline
           & Rel. & $5.455(0.012)$ & $5.736(0.013)$ & $0.282(0.024)$ & $5.782(0.023)$ & $5.892(0.022)$ & $2.395(0.001)$ \\
$^{208}$Pb & Non. & $5.447(0.014)$ & $5.607(0.020)$ & $0.161(0.033)$ & $5.774(0.023)$ & $5.885(0.022)$ & $2.283(0.005)$ \\ \cline{2-8}
           & Exp. & \multicolumn{3}{c|}{$R_c=5.503(0.014)$} & \multicolumn{3}{c|}{$Q_c=5.851(0.022)$} \\ \hline
\end{tabular}
}
\endgroup
\caption{
The results of the least square analysis. The numbers in the parentheses denote
the error coming from experiment. They are obtained, neglecting the standard deviation of the
calculated values from the least square line. All the numbers are given in units of fm.
}
\end{table}

\begin{table}
\begingroup
\renewcommand{\arraystretch}{1.2}
\hspace*{-1cm}%
{\setlength{\tabcolsep}{4pt}
\begin{tabular}{|c|l|l|l|l|l|l|l|} \hline
          &      &
\multicolumn{1}{c|}{$R_p$} &
\multicolumn{1}{c|}{$R_n$} &
\multicolumn{1}{c|}{$\delta R$} &
\multicolumn{1}{c|}{$Q_p$} &
\multicolumn{1}{c|}{$Q_{cp}$} & 
\multicolumn{1}{c|}{$Q_{cn}$} \\ \hline
           & Rel. & $3.348(0.003)$ & $3.297(0.006)$ & $-0.050(0.009)$ & $3.635(0.015)$ & $3.785(0.014)$ & $1.512(0.004)$ \\
$^{40}$Ca  & Non. & $3.348(0.009)$ & $3.304(0.011)$ & $-0.044(0.020)$ & $3.628(0.015)$ & $3.782(0.014)$ & $1.465(0.004)$ \\ \cline{2-8}
           & Exp. & \multicolumn{3}{c|}{$R_c=3.450(0.010)$} & \multicolumn{3}{c|}{$Q_c=3.761(0.014)$} \\ \hline
           & Rel. & $3.378(0.005)$ & $3.597(0.021)$  & $0.220(0.026)$ & $3.643(0.014)$ & $3.796(0.012)$ & $1.897(0.006)$ \\
$^{48}$Ca  & Non. & $3.372(0.009)$ & $3.492(0.028)$  & $0.121(0.036)$ & $3.629(0.014)$ & $3.786(0.013)$ & $1.811(0.009)$ \\ \cline{2-8}
           & Exp. & \multicolumn{3}{c|}{$R_c=3.451(0.009)$} & \multicolumn{3}{c|}{$Q_c=3.736(0.012)$} \\ \hline
           & Rel. & $5.454(0.013)$ & $5.728(0.057)$ & $0.275(0.070)$ & $5.783(0.023)$ & $5.892(0.022)$ & $2.395(0.008)$ \\
$^{208}$Pb & Non. & $5.447(0.014)$ & $5.609(0.054)$ & $0.162(0.068)$ & $5.774(0.023)$ & $5.885(0.022)$ & $2.283(0.009)$ \\ \cline{2-8}
           & Exp. & \multicolumn{3}{c|}{$R_c=5.503(0.014)$} & \multicolumn{3}{c|}{$Q_c=5.851(0.022)$} \\ \hline
\end{tabular}
}
\endgroup
\caption{The results of the least square analysis. The numbers in the parentheses denote
the error which is obtained taking account of the experimental error and the standard deviation of the
calculated values from the least square line.  All the numbers are given in units of fm. }
\end{table}

In $^{40}$Ca, most of the non-relativistic models predict the larger values
of $R_c$ than those in the relativistic models, and overestimate
its experimental value. All the calculated values in both models, however,
are almost on the same LSL between $R^2_p$ and $R^2_c$. 
As a result, the values of $R_p$ are determined to be almost the same
in the two frameworks, as shown in Table 6 and 7. 
The value of $R_n$ is also estimated to be almost the same in the two models,
but to be smaller by $0.04\sim$ 0.05 fm than that of $R_p$, as expected from the
Coulomb force.
The difference between the values of $Q_{cn}$ in the two models
is mainly due to the contribution
from the spin-orbit density which is enhanced more in the relativistic models
than in the non-relativistic ones.
The same enhancement in $Q_{cn}$ is also seen in $^{48}$Ca and $^{208}$Pb
in Table 6 and 7.

In $^{48}$Ca, on the one hand, $R^2_c$ is overestimated by the non-relativistic models
in the same way as in  $^{40}$Ca. Nevertheless, its LSL with $R^2_p$ is almost the same
as that of the relativistic models, although there is a small difference between them
owing to the spin-orbit density corrections. The LSL's yield
$R_p\approx 3.37\sim 3.38$ fm, which is larger by $0.02 \sim$ 0.03 fm than that of $^{40}$Ca.
This difference is cancelled by the negative contribution from
the neutron charge density to reproduce almost the same experimental
value of $R^2_c$ in $^{40}$Ca and $^{48}$Ca.
On the other hand, the values of $R^2_n$ in $^{48}$Ca
evaluated in the relativistic and non-relativistic models are distributed
in the same region around 13.0 fm$^2$,
as shown in Figure \ref{RnRc-Ca48}, \ref{RnQc-Ca48} and \ref{RnQcn-Ca48}.
The LSA, however, yields a larger value of $R^2_n$ for the
relativistic models by $\sim 0.1$ fm than that for the non-relativistic models.
As a result, the value of $\delta R$ is larger by $\sim 0.1$ fm
in the relativistic models than in the non-relativistic ones.

The values of $Q_{cp}$ is larger than those of $Q_{cn}$ in both $^{40}$Ca
and $^{48}$Ca in Table 6 and 7. The values of $Q_{cp}$, however, are almost the same
in the two nuclei, while the value of $Q_{cn}$ in $^{48}$Ca is lager than
that in $^{40}$Ca. These results explain the fact that the experimental value
of $Q^4_c$ of $^{48}$Ca is smaller than that of ${^{40}}$Ca \cite{emrich},
as indicated in Figure \ref{RpQc-Ca40} and \ref{RpQc-Ca48},
since $Q^4_{cn}$ provides a negative contribution to $Q^4_c$.
The negative contribution from the neutrons is also expected  to explain
the fact that the value of the sixth-order moment is smaller in $^{48}$Ca
than in $^{40}$Ca\cite{kss}.
The investigation of the sixth-order moments may yield more detailed information
not only on $Q_{cp},Q_{cn}$ and $Q_p$, but also on the fourth-order moment $Q_n$
of the neutron density which has not been explored in the present paper.
 
In $^{208}$Pb, like the relativistic models,
some of the non-relativistic models predict almost the experimental value
of $R^2_c$, in contrast to the cases in $^{40}$Ca and $^{48}$Ca.
This result affects the distribution of the predicted values of $R^2_n$
in the $R^2_n-R^2_c$, $R^2_n-Q^4_c$ and $R^2_n-Q^4_{cn}$ plane
in the non-relativistic models.
Some of the predicted values are on the intersection regions between
the LSL's and those for the experimental values
with the errors of $R^2_c$ and $Q^4_c$. 
The estimated value of $R_n$, however, is smaller by $\sim$ 0.1 fm
in the non-relativistic models than that in the relativistic models,
just as in $^{48}$Ca.
The difference by 0.1 fm is shown to play an essential role
in the discussions on the size of the neutron star\cite{thi}.
It is under investigation what causes the difference between $R_n$ or $\delta R$
in the relativistic and non-relativistic frameworks,
in addition to the relativistic corrections to $R_n$\,\cite{ksp}.

Finally, three general comments are added.
First,
in the present paper, all the numbers have been kept up to the third decimal
place, according to the experimental values\cite{emrich}. We note that
if models with different parameterizations of the nuclear interactions
are added, or other single-nucleon form factors are used in the analysis,
the number of the second decimal place would be changed.
Furthermore, ambiguity of the relativistic corrections to
the non-relativistic models, which stems from the inconsistency between them,
may change the number of the second decimal place.
Neglecting the exchange term of the Coulomb force in the non-relativistic
models as in the relativistic cases may affect the number in the same place. 
The general conclusions derived by the present LSA, however,
are expected to be unchanged. When new phenomenological interactions
are explored, the obtained various LSL's 
will provide a convincing guide to search their new parameters.

Second,
the detailed investigations on $Q_c$ together with $R_c$ in this paper
may be useful for understanding the parity-violating electron scattering
already performed at $q=0.475 {\rm fm}^{-1}$\cite{abra}, where both moments contribute
to its cross section\cite{ksmsr}.
The present analyses are also expected to play a complementary role
in the study of the neutron distribution under planning\cite{thi}.

Third,
the results obtained in the present paper bring a good prospect in the study of
unstable nuclei. It is one of the most important
problems to explore not only the change of the proton density,
but also that of the neutron density from those in stable nuclei,
since the stability of such nuclei is dominated
by the structure of the neutron distribution.
As the contributions from the neutron density to the charge density
are expected to increase in unstable nuclei,
both proton and neutron distributions would be investigated more clearly 
through electromagnetic interaction with less ambiguity
than through other experimental approach\cite{thi}.
This fact implies that the new electron scattering facilities
in the world\cite{suda,tsukada} make the forthcoming study
of unstable nuclei more efficient and stimulating.

\section*{Acknowledgment}
The authors would like to thank Professor M. Wakasugi, Professor T. Tamae
and Dr. K. Tsukada for useful discussions.

\vspace{7mm}
\noindent
{{\Large {\bf Appendix}}

\appendix

\section{Least squares analysis}

\renewcommand{\theequation}{A.\arabic{equation}}
\setcounter{equation}{0}
The least squares analysis(LSA) of the set ($x_i, y_i\,, i=1, 2,\cdots n$)
provides for the linear relationship between the two quantities $x$ and $y$ as
\begin{equation}
\hat{y}=ax+b,\label{lsl}
\end{equation}
where the  coefficients $a$ and $b$ are given by
\begin{align}
  a&=\frac{\avr{xy}-\avr{x}\avr{y}}{\avr{x^2}-\avr{x}^2}\,,\\
  b&=\avr{y}-a\avr{x}=\frac{\avr{x}^2\avr{y}-\avr{x}\avr{xy}}{\avr{x^2}-\avr{x}^2}\,.
\end{align}
Here $\avr{x}$, etc. denote the mean values of the elements in the set.
The slope of the least square line(LSL) in Eq.(A.1) is expressed in terms of
the correlation coefficient $r_{xy}$ as
\begin{equation}
a=\frac{\Delta y}{\Delta x}r_{xy}\,, \quad r_{xy}=\frac{\avr{xy}-\avr{x}\avr{y}}{\Delta x\Delta y}\,
\end{equation}
with
\begin{equation}
\Delta x=\sqrt{\avr{x^2}-\avr{x}^2}\,,\quad \Delta y=\sqrt{\avr{y^2}-\avr{y}^2}.
\end{equation}
The standard deviation of the elements from Eq.(A.1), $\sigma$, is defined by
\begin{equation}
n\sigma^2=\sum^n_{i=1}(y_i-ax_i-b)^2.
\end{equation}
The relationship between $r_{xy}$ and $\sigma$ is expressed as
\begin{equation}
r^2_{xy}=1-\frac{\sigma^2}{(\Delta y)^2}=\frac{a^2(\Delta x)^2}{a^2(\Delta x)^2+\sigma^2},
\end{equation}
which shows that $r_{xy}=1$ for $\sigma=0$ and it decreases with increasing $\sigma$,
or with decreasing $\Delta y$,
and that $r_{xy}\approx 0$, when $\sigma\approx \Delta y$, or
$a\Delta x \ll \sigma$. The closer to 1
the value of $r_{xy}$ is, the higher the validity of the LSA is relatively.

\section{Correlation coefficient}
\renewcommand{\theequation}{B.\arabic{equation}}
\setcounter{equation}{0}
In the text, the LSA has not been applied to the analysis of the relationship between
$Q^4_{cn}$ and $Q^4_c$ in the relativistic models, since the distribution of their elements
in the $Q^4_{cn}$ and $Q^4_c$ plane seems not to be appropriate for the LSA.
This fact is explored numerically in terms of the correlation coefficients as follows,

The three kinds of the LSA have been performed between $R^2_p$, $Q^4_c$ and $Q^4_{cp}$
for the proton density in the text. The obtained LSL are described as, 
\begin{equation}
\hat{q}_x=a_{xq}x+b_{xq}\, ,\quad \hat{q}_p=a_{pq}p+b_{pq}\,,\quad
\hat{p}=a_{xp}x+b_{xp},
\end{equation}
where $x$, $q$ and $ p$ denote $R^2_p$, $Q^4_c$ and $Q^4_{cp}$, respectively.
If $(x_i, q_i)$ and $(p_i, q_i)$ are on the LSL and $\hat{q}_{x,i}=\hat{q}_{p,i}=q_i$
and $\hat{p}_i=p_i$, then Eq.(B.1) yield 
\begin{equation}
q_i=a_{pq}(a_{xp}x_i+b_{xp})+b_{pq}=a_{xp}a_{pq}x_i+a_{pq}b_{xp}+b_{pq}=a_{xp}x_i+b_{xp}.
\end{equation}
Since the above equation holds for any $i$, the slopes and intercepts of Eq.(B.1)
should satisfy
\begin{equation}
a_{xp}a_{pq}-a_{xq}=0\,,\quad b_{xp}a_{pq}+b_{pq}-b_{xq}=0.
\end{equation}
Using the correlation coefficients, the first equation in Eq.(B.3) is expressed as
\begin{equation}
r_{xp}r_{pq}/r_{xq}=1,
\end{equation}
which holds when $r_{xp}=r_{pq}=r_{xq}=1$.
According to Eq.(A.7), the above equation is used as a guide of the validity for LSA,
together with the values of the correlation coefficients themselves,
since actually the elements of the set($x_i, q_i, p_i$) with $\sigma\neq 0$
are not on the LSL.
In the mean field models used in the text, they are given, for example, in $^{208}$Pb as
\begin{equation}
r_{xq}=0.9967\,,\quad r_{xp}=0.9960\,,\quad r_{pq}=0.9999\,, \quad
 r_{xp}r_{pq}/r_{xq}=0.9993
\end{equation}
in the relativistic models, while in the non-relativistic models,
 \begin{equation}
r_{xq}=0.9863\,,\quad r_{xp}=0.9859\,,\quad r_{pq}=1.0000\,, \quad
 r_{xp}r_{pq}/r_{xq}=0.9995.
\end{equation} 
Thus, all the values of the correlation coefficients are nearly equal to 1 and
the relationship $a_{xp}a_{pq}/a_{xq}=1$ holds almost exactly in both models,
reflecting $\sigma\approx0$. For other nuclei also, similar results have been obtained.

The same analysis is performed for the neutron density as for the proton density. 
In this case, $x$ should be read as $R^2_n$, and $p$ as $n=Q^4_{cn}$.
The calculated values of the correlation coefficients are given, for example, for $^{208}$Pb as
\begin{equation}
r_{xq}=0.8039\,,\quad r_{xn}=0.3689\,,\quad r_{nq}=0.1910\,, \quad
 r_{xn}r_{nq}/r_{xq}=0.0877
\end{equation}
in the relativistic models, while in the non-relativistic models, they are
\begin{equation}
r_{xq}=0.8780\,,\quad r_{xn}=0.9951\,,\quad r_{nq}=0.8730\,, \quad
 r_{xn}r_{nq}/r_{xq}=0.9895.
\end{equation}
It is seen that the value of $r_{nq}$ is small, compared with others in the above
two equations. 
The reason of the small value of $r_{nq}$ is understood in Eq.(A.7).
As seen in Figure \ref{QcnQc-Ca40} for $^{40}$Ca, the value of $a\Delta x$ is comparable
with that of $\sigma$ in Table 3.

The small value of $r_{nq}$ in the relativistic models causes another problem.
It violates the definition of the relationship,
$Q^4_c=Q^4_{cp}-Q^4_{cn}$ on the LSL's.
The LSL between $Q^4_{cp}$ and $Q^4_c$ and that
between $Q^4_{cn}$ and $Q^4_c$ are written as
\begin{equation}
\hat{q}_p=a_{pq}p+b_{pq}\,,\quad \hat{q}_n=a_{nq}n+b_{nq}\,.
\end{equation}
When the values of $R^2_p=p_e$ and $R^2_n=n_e$ are determined by the intersection
points of the above LSL and the experimental value of $Q^4_c=\hat{q}_p=\hat{q}_n=q_e$, 
Eq.(B.9) provides
\begin{equation}
q_e=a_{pq}p_e+b_{pq}=a_{nq}n_e+b_{nq},
 \end{equation}
 which yields
 \begin{equation}
  p_e-n_e=\left(\frac{1}{a_{pq}}-\frac{1}{a_{nq}}\right)q_e
   +\left(\frac{b_{nq}}{a_{nq}}-\frac{b_{pq}}{a_{pq}}\right)
\end{equation}
The first term of the right-hand side is expressed
in terms of the correlation coefficients as
\begin{equation}
\frac{1}{a_{pq}} - \frac{1}{a_{nq}}=\lambda+1\,,\quad \lambda
 =\frac{r^2_{pq} - 1}{r^2_{pq} - a_{pq}}=\frac{r^2_{nq} - 1}{r^2_{nq} + a_{nq}}
\end{equation}
The calculated values of the elements($n_i,p_i,q_i$) satisfy the definition, $n_i=p_i-n_i$,
so that they are written as
\begin{equation}
q_i=a_{pq}p_i+b_{pq}+\epsilon^p_i=a_{nq}n_i+b_{nq}+\epsilon^n_i\,,\quad \avr{q}=\avr{p}-\avr{n}\,,
\end{equation}
where $\epsilon^p_i$ and $\epsilon^n_i$ represent the deviation from LSL with
$\avr{\epsilon^p}=\avr{\epsilon^n}=0$.
The above equation gives
\begin{equation}
\left(\frac{b_{pq}}{a_{pq}}-\frac{b_{nq}}{a_{nq}}\right)=\lambda \avr{q}.
\end{equation}
Using Eq.(12) and (B.14), Eq.(B.11) is finally described as
\begin{equation}
p_e-n_e=q_e+\lambda(q_e - \avr{q}).
\end{equation}
This shows that the relationship between $Q^4_{cp}$, and $Q^4_{cn}$ and $Q^4_c$ by the definition
is violated, unless $\lambda= 0$, or $\avr{q}=q_e$.
In the relativistic models for $^{208}$Pb, the value of $\lambda$ is -0.0687
with $r_{nq}$ in Eq.(B.7) and that of $\avr{q}$ is 1170.2928 fm$^4$,
for $\hat{q}_e=1171.981$ fm$^4$. In the non-relativistic models, those values are
given as $\lambda=-0.0038$ and $\avr{q}=1173.9587$ fm$^4$.

Thus, it is reasonable from a numerical point of view also that the LSA between $Q^4_{cn}$
and $Q^4_c$ has been excluded in the present analysis. 

In Eq.(B.7) and (B.8), the value of $r_{xn}$ in the relativistic models is also rather small,
compared with others. The small value is understood, according to Eq.(A.7). In this case,
$\Delta y$ is small, as seen in Figure \ref{RnQcn-Pb208}, since the calculated values of $Q^4_{cn}$
are concentrated in the narrow region in the same way as in Figure \ref{QcnQc-Ca40}. 
In the present analysis, the results of LSA on the relationship between $R^2_{cn}$
and $Q^4_{cn}$ have been positively taken into account, because of the small value of $\sigma$
in Table 3, 4 and 5. In the final results, however, the only lower bound of the common accepted region
for $R^2_n$ in $^{208}$Pb is determined by this relation in neglecting $\sigma$,
as mentioned in the text. All other common accepted regions in the relativistic models are 
within the regions determined by the relationship between $R^2_n$ and $Q^4_{cn}$.

\section{The accepted region}
\renewcommand{\theequation}{C.\arabic{equation}}
\setcounter{equation}{0}

The standard deviation $\sigma$ has been taken into account in the following way.

The intersection point of the LSL in Eq.(A.1) with the  line of the experimental value
$\hat{y}=y_e$ determines the value $x_e$ of $x$. In denoting the experimental error
by $\delta y_e$, the intersection point of the LSL with the line
of $\hat{y}=y_e\pm\delta y_e$ provides the value of $x$ as $x_e\pm\delta x$
with $\delta x = \delta y_e/a$. In neglecting $\sigma$, then the accepted region
of $x_e\pm\delta x$ is expressed as $x_e-\delta x\sim x_e+\delta x$, or $x_e(\delta x)$ in the text.

When the standard deviation of LSL is taken into account, the LSL is replaced by
$\hat{y}_{\pm}=ax +b \mp \sigma$. The intersection points with the lines
of $\hat{y}=y_e\pm\delta y_e$ yield the accepted region $\mathcal{R}$ to be
\begin{equation}
\mathcal{R}=x_e\pm\delta x\,, \quad \delta x = (\delta y_e+\sigma)/a.
\end{equation}
If the LSL between $R^2_p(x)$ and $R^2_c(d)$ is described as
\begin{equation}
\hat{d}=a_{xd}x+b_{xd},
\end{equation}
then the accepted region $\mathcal{R}_d$ of $x$ is given by
\begin{equation}
\mathcal{R}_d=x_{ed}\pm\delta x_{xd}\,, \quad \delta x_{xd}=(\delta d_e +\sigma_{xd})/a_{xd},
\end{equation}
where $x_{ed}$ is determined by Eq.(C.2) with the experimental value $d_e=\hat{d} $,
and its error is denoted by  $\delta d_{xd}$ and the standard deviation of LSL by $\sigma_{xd}$.
In the case of the relationship between $R^2_n$ and $R^2_c$,
$x$ in the above two equations is replaced by $R^2_n$.

For the analyses of $R^2_p(x)$ and $Q^4_c(q)$ and of $R^2_n(x)$ and $Q^4_c(q)$, the accepted region
$\mathcal{R}_q$ is determined in the same way as in Eq.(C.3),
\begin{equation}
 \mathcal{R}_q=x_{eq}\pm\delta x_{xq}\,, \quad \delta x_{xq}=(\delta q_e+\sigma_{xq} )/a_{xq},
\end{equation}
Here, $x_{eq}$ stands for the intersection point of the LSL $\hat{q}_x$ in Eq.(B.1)
with the line of the experimental value of $\hat{q}_x=q_e$, and $\delta q_e$ denotes the error
 of the experimental value of $q_e$ and $\sigma_{xq}$ the standard deviation of the LSL.

In the case of $R^2_p(x)$ and $Q^4_{cp}(p)$, the analysis has been performed by the two steps.
First, the relationship between $Q^4_{cp}$ and $Q^4_c(q)$ is analyzed in order to determine
the {\it pseudo} experimental value of $Q^4_{cp}$ with the error $\delta p_e$,
which is given by
\begin{equation}
\delta p_e =(\delta q_e + \sigma_{pq})/a_{pq},
\end{equation}
with the standard deviation $\sigma_{pq}$ of LSL of the first equation in Eq.(B.9).
Next, the accepted region of $R^2_p$ is estimated from the relationship
between $R^2_p$ and $Q^4_{cp}$, using the pseudo experimental value. 
Then, the accepted region $\mathcal{R}_{pq}$ of $x$ is given by
\begin{equation}
\mathcal{R}_{pq}=x_{ep}\pm\delta x_{xp}\,,\quad \delta x_{xp}=(\delta p_e + \sigma_{xp})/a_{xp}
 =(\delta q_e+\sigma_{pq}+a_{pq}\sigma_{xp})/(a_{xp}a_{pq}),
 \end{equation}
where $x_{ep}$ stands for the intersection point of the LSL of the third equation in Eq.(B.1)
with $\hat{p}=p_e$, and $\sigma_{xp}$ the standard deviation of the LSL.
If the standard deviations are negligible, then $\mathcal{R}_q=\mathcal{R}_{pq}$, because of
Eq.(B.3), as numerically seen in the text.

The relationship between $R^2_n$ and $Q^4_{cn}$ is also explored by the two steps, but in a different
way form those for $R^2_p$.
First, the error $\delta n_e$ of the {\it pseudo} experimental value of $Q^4_{cn}$  
is determined through $Q^4_c$  and $Q^4_{cp}$ as 
\begin{equation}
\delta n_e = \delta p_e -\delta q_e =\left((1-a_{pq})\delta q_e+\sigma_{pq}\right)/a_{pq}.
 \end{equation}
Second, the analysis of the relationship between $R^2_n$ and $Q^4_{cn}$ gives the accepted 
region $\mathcal{R}_{nq}$ of $R^2_n$ as
\begin{align}
\mathcal{R}_{nq}=x_{en}\pm &\delta x_{xn}\,, \nonumber\\
 &\delta x_{xn}=(\delta n_e +\sigma_{xn})/a_{xn}
 =\left((1-a_{pq})\delta q_e+\sigma_{pq}+a_{pq}\sigma_{xn}\right)/(a_{xn}a_{pq}),
 \end{align}
Here, the LSL of the relationship between $R^2_n(x)$ and $Q^4_{cn(n)}$ is described as
\begin{equation}
 \hat{n}=a_{xn}x+b_{xn}
 \end{equation}
 with the standard deviation $\sigma_{xn}$, and its the intersection point with the pseudo
 experimental value $\hat{n}=n_e$ is denoted by $x_{en}$.
If all the standard deviations are neglected in Eq.(C.8), then $\mathcal{R}_{nq}=\mathcal{R}_q$
holds for the neutrons, because of Eq.(B.11) and of the equation replacing $p$ with $n$ in Eq.(B.3).

 \end{document}